\documentclass[11pt]{article}
\pdfoutput=1
\usepackage{fullpage}
\usepackage{xcolor}
\usepackage{cite}
\usepackage{graphicx}
\usepackage{tensor}
\usepackage{amsmath}
\usepackage{amssymb}
\usepackage{subcaption}
\usepackage[utf8x]{inputenc} 
\title{}
\date{}

\def\beq{\begin{equation}}
\def\eeq{\end{equation}}
\allowdisplaybreaks
\begin{document}
\bibliographystyle{utphys}

\newcommand{\be}{\begin{equation}}
\newcommand{\ee}{\end{equation}}
\newcommand\n[1]{\textcolor{red}{(#1)}} 
\newcommand{\diff}{\mathop{}\!\mathrm{d}}
\newcommand{\lb}{\left}
\newcommand{\rb}{\right}
\newcommand{\f}{\frac}
\newcommand{\pd}{\partial}
\newcommand{\tr}{\text{tr}}
\newcommand{\fdiff}{\mathcal{D}}
\newcommand{\im}{\text{im}}
\let\caron\v
\renewcommand{\v}{\mathbf}
\newcommand{\T}{\tensor}
\newcommand{\R}{\mathbb{R}}
\newcommand{\C}{\mathbb{C}}
\newcommand{\Z}{\mathbb{Z}}
\newcommand{\msbar}{\ensuremath{\overline{\text{MS}}}}
\newcommand{\DIS}{\ensuremath{\text{DIS}}}
\newcommand{\abar}{\ensuremath{\bar{\alpha}_S}}
\newcommand{\bb}{\ensuremath{\bar{\beta}_0}}
\newcommand{\rc}{\ensuremath{r_{\text{cut}}}}
\newcommand{\Nd}{\ensuremath{N_{\text{d.o.f.}}}}
\newcommand{\red}[1]{{\color{red} #1}}
\setlength{\parindent}{0pt}

\titlepage
\begin{flushright}
QMUL-PH-22-15\\
\end{flushright}

\vspace*{0.5cm}

\begin{center}
{\bf \Large Non-perturbative aspects of the self-dual double copy}

\vspace*{1cm} 
\textsc{Kymani Armstrong-Williams\footnote{k.t.k.armstrong-williams@qmul.ac.uk},
Chris D. White\footnote{christopher.white@qmul.ac.uk},
  and Sam Wikeley\footnote{s.wikeley@qmul.ac.uk}} \\

\vspace*{0.5cm} Centre for Theoretical Physics, Department of
Physics and Astronomy, \\
Queen Mary University of London, 327 Mile End
Road, London E1 4NS, UK\\

\end{center}

\vspace*{0.5cm}

\begin{abstract}
  The double copy is by now a firmly-established correspondence
  between amplitudes and classical solutions in biadjoint scalar,
  gauge and gravity theories. To date, no strongly coupled examples of
  the double copy in four dimensions have been found, and previous
  attempts based on exact non-linear solutions of biadjoint theory in
  Lorentzian signature have failed. In this paper, we instead look for
  biadjoint solutions in Euclidean signature, which may be relatable
  to Yang-Mills or gravitational instantons. We show that spherically
  symmetric power-like Euclidean solutions do not exist in precisely
  four spacetime dimensions. The explanation for why this is the case
  turns out to involve the Eguchi-Hanson instanton, whose single copy
  structure is found to be more complicated (and interesting) than
  previously thought. We provide a more general prescription for
  double-copying instantons, and explain how our results provide a
  higher-dimensional complement to a recently presented
  non-perturbative double copy of exact solutions in two spacetime
  dimensions. In doing so, we demonstrate how the replacement of
  colour by kinematic Lie algebras operates at the level of exact
  classical solutions.
\end{abstract}

\vspace*{0.5cm}

\section{Introduction}
\label{sec:intro}

\begin{figure}
  \begin{center}
    \scalebox{0.8}{\includegraphics{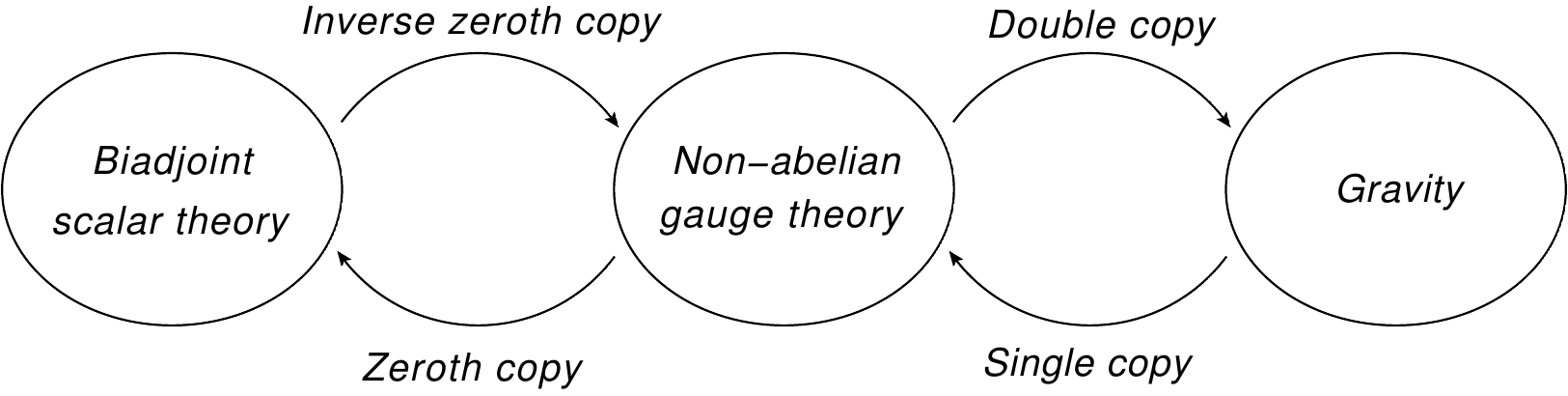}}
    \caption{Correspondences between various types of field theory.}
    \label{fig:theories}
  \end{center}
\end{figure}

In recent years, a correspondence known as the {\it double copy} has
received a great deal of attention. Originating in the study of field
theory scattering amplitudes~\cite{Bern:2010ue,Bern:2010yg}, and inspired by earlier work in string theory~\cite{Kawai:1985xq}, it
states that quantities in non-abelian gauge theories can be
straightforwardly mapped to counterparts in gravity
theories. Similarly, one can take objects in gauge theory, and
translate them to a so-called {\it biadjoint scalar theory}, in which
a scalar field carries two types of colour charge. This is known as
the {\it zeroth copy}, and it is common to depict the ladder of
correspondences between these different theories as in
figure~\ref{fig:theories}, although this is itself a subset of a much
wider web of theories (see e.g.~\cite{Bern:2019prr} for a review).\\

Much of the work in recent years has focused on how generally we are
to interpret the scheme of figure~\ref{fig:theories}. We unfortunately
lack a complete understanding of the double copy at the level of
e.g. Lagrangians or equations of motion (although see
refs.~\cite{Bern:2010yg,Monteiro:2011pc,Cheung:2016say,Cheung:2017kzx,Tolotti:2013caa,Borsten:2020xbt,Borsten:2020zgj,Borsten:2021hua,Beneke:2021ilf}
for interesting developments), so that it is not known whether
figure~\ref{fig:theories} applies to the complete theories. If it
does, it suggests a profound and previously hidden commonality between
our theories of nature, that traditional ways of thinking have
obscured. Furthermore, there is a practical way to test how generally
figure~\ref{fig:theories} applies. Namely, to take different objects
in each of the theories, and to find rules for matching these up that
can be seen to generalise the original double copy for scattering
amplitudes. The first such work in this regard was
ref.~\cite{Monteiro:2014cda}, which extended the double and zeroth
copies to a special family of exact classical solutions of the
relevant theories: those associated with {\it Kerr-Schild metrics} in
General Relativity (see also
refs.~\cite{Didenko:2008va,Didenko:2009td} for related work in a
different context). Follow-up work has considered the implications for
specific solutions, including in different numbers of spacetime
dimension~\cite{Luna:2015paa,Ridgway:2015fdl,Bahjat-Abbas:2017htu,Berman:2018hwd,Carrillo-Gonzalez:2017iyj,CarrilloGonzalez:2019gof,Bah:2019sda,Alkac:2021seh}. A
second exact double copy procedure is the {\it Weyl double copy}
introduced in ref.~\cite{Luna:2018dpt}, and studied further in
refs.~\cite{Sabharwal:2019ngs,Alawadhi:2020jrv,Godazgar:2020zbv,White:2020sfn,Monteiro:2020plf,Chacon:2021wbr,Chacon:2021hfe,Chacon:2021lox,Godazgar:2021iae,Monteiro:2021ztt}.
This is complementary to the Kerr-Schild double copy in that it relies
on the spinorial formulation of field theory rather than the tensorial
approach. Nevertheless, it agrees where overlap exists. In addition to
exact solutions, one may double-copy classical solutions
order-by-order in perturbation theory. For a variety of classical
double copy approaches, see
e.g. refs.~\cite{Elor:2020nqe,Farnsworth:2021wvs,Anastasiou:2014qba,LopesCardoso:2018xes,Anastasiou:2018rdx,Luna:2020adi,Borsten:2020xbt,Borsten:2020zgj,Borsten:2021hua,Goldberger:2017frp,Goldberger:2017vcg,Goldberger:2017ogt,Goldberger:2019xef,Goldberger:2016iau,Prabhu:2020avf,Luna:2016hge,Luna:2017dtq,Cheung:2016prv,Cheung:2021zvb,Cheung:2022vnd,Cheung:2022mix,Campiglia:2021srh}.\\

In all of the above double copies, solutions of {\it linearised}
biadjoint scalar theory play a crucial role. They emerge as the
denominators in scattering amplitudes, or as intermediate quantities
in classical double copies. To date, there has been no concrete
realisation of figure~\ref{fig:theories} for genuinely non-perturbative
/ strongly coupled solutions in four dimensions, involving fully
non-linear solutions of each theory, including the biadjoint scalar
case. A number of ideas have been proposed for exploring
non-perturbative aspects of the double copy, including the study of how
symmetries in different theories can be
mapped~\cite{Monteiro:2011pc,Borsten:2021hua,Alawadhi:2019urr,Banerjee:2019saj,Huang:2019cja},
or non-trivial geometric / topological
properties~\cite{Berman:2018hwd,Alfonsi:2020lub,Alawadhi:2021uie}. Recently,
a proposal for a non-perturbative double copy in two spacetime
dimensions has been made~\cite{Cheung:2022mix}, and we return to this
below. First, however, we continue a programme of work initiated in
ref.~\cite{White:2016jzc}, which found exact non-linear solutions of
biadjoint scalar theory. Further solutions were found in
refs.~\cite{DeSmet:2017rve,Bahjat-Abbas:2018vgo}, and the hope is that
by assembling a catalogue of such solutions, it might be possible to identify their counterparts in gauge or gravity theory, thus
providing a non-perturbative realisation of the double copy and related
correspondences.\\

The simplest non-linear solutions of biadjoint theory consist of
spherically-symmetric monopole-like objects, involving a singularity
at the origin. It is then natural to propose that these might be
associated with singular monopoles in Yang-Mills theory, namely the
{\it Wu-Yang monopoles} of ref.~\cite{Wu:1967vp}. Indeed, this
conjecture was tentatively made in ref.~\cite{White:2016jzc}, but
subsequent work has shown that it cannot be
true~\cite{Bahjat-Abbas:2020cyb,Alfonsi:2020lub}. The Wu-Yang monopole
turns out to be related (by a singular gauge transformation) to a
non-abelian version of the well-known Dirac magnetic monopole, whose
double and zeroth copies are already known: in gravity, it
corresponds to a so-called {\it NUT charge}~\cite{Taub,NUT}, as first
shown in ref.~\cite{Luna:2015paa}. Thus, there is no room left for the
``biadjoint monopole'' to correspond to an obvious gauge theory
solution, and it remains unclear how to proceed.\\

Given the above uncertainty regarding monopole solutions -- which are
in standard Lorentzian signature -- we take here a different approach. Besides monopoles, some of the most well-known
non-perturbative solutions of gauge and gravity theories are {\it
  instantons}, which are solutions of the field equations in {\it
  Euclidean} signature. There is thus a clear motivation for trying to
find non-linear solutions of Euclidean biadjoint scalar theory, in the hope that they may be relatable to known instantons. That this should
be possible is further motivated by the fact that instanton solutions,
as commonly referred to in both Yang-Mills (YM) theory and gravity,
are (anti)-self-dual. It is known, at least in principle, that the
self-dual sectors of YM and gravity can be written as manifest double copies
of each other~\cite{Monteiro:2011pc}, where the zeroth copy to
biadjoint theory also takes a simple form. Generalisations of this
idea to more exotic theories are also known~\cite{Chacon:2020fmr},
although it is not clear what the precise implications are for
classical solutions rather than scattering amplitudes.\\

In this paper, we will start by trying to find simple power-like
spherically symmetric solutions of Euclidean biadjoint scalar field
theory, analogous to the Lorentzian solutions found in
ref.~\cite{White:2016jzc}. Curiously, we will see that whilst non-linear
solutions do indeed exist in $d\neq 4$ spacetime dimensions, they are
entirely absent in $d=4$. This may at first seem surprising, but
is in fact easily explainable: the power-like
solutions one obtains for general dimensions solve the {\it linearised}
biadjoint equation in $d=4$, and thus cannot be solutions of the
non-linear equation. Based on previous
insights~\cite{Berman:2018hwd,Luna:2018dpt}, we are able to identify
these linear solutions with the zeroth copy of the Eguchi-Hanson
(gravitational)
instanton~\cite{Eguchi:1978xp,Eguchi:1978gw,Eguchi:1979yx}. However,
in doing so, we find that the single copy of the Eguchi-Hanson
instanton is more intricate than previously thought: one may provide a
fully non-abelian single copy of this solution, in addition to its
previously understood abelian counterpart. This mirrors the situation
found for monopoles in
refs.~\cite{Bahjat-Abbas:2020cyb,Alfonsi:2020lub}, where either an
abelian or non-abelian monopole are found to double copy to the same
gravity solution (a NUT charge). We will also be able to write a more
general ansatz for double-copying instantons than has previously been
used. However, ultimately we find that it applies only to those
gauge or gravity solutions which linearise the equations of
motion.\\

As mentioned above, ref.~\cite{Cheung:2022mix} recently proposed a
non-perturbative double copy procedure for exact solutions in a
variety of theories in two spacetime dimensions. The role of the
gravity theory is played by {\it Special Galileon (SG) theory}, and
that of the gauge theory by so-called {\it Zakharov-Mikhailov (ZM)
  theory}. The equations of motion for these theories, which we write explicitly in section~\ref{sec:Cheung}, bear a strong resemblance to those of self-dual
  Yang-Mills theory and gravity in four spacetime dimensions. This will
allow us to interpret the results of ref.~\cite{Cheung:2022mix} as a close counterpart of the known four-dimensional self-dual double
copy. Conversely, the ideas of ref.~\cite{Cheung:2022mix} will also
clarify aspects of our four-dimensional results. In particular, the
original double copy for scattering amplitudes relies on a phenomenon
known as {\it BCJ duality}~\cite{Bern:2008qj}, which states that there
is a {\it kinematic algebra} underlying gauge theory amplitudes,
mirroring the Lie algebra describing the colour degrees of
freedom. The gravity theory has no colour algebra, but instead has two
copies of the kinematic algebra. The nature of this algebra has
remained mysterious in general, but it can be made explicit at the level of equations of motion in the (anti-)self-dual sector, where it is known
to correspond to a certain infinite-dimensional Lie algebra of
area-preserving diffeomorphisms~\cite{Monteiro:2011pc}. Until now, it has remained unclear
how this algebra relates to properties of (exact) classical solutions,
but ref.~\cite{Cheung:2022mix} provides an answer to this question in
two spacetime dimensions. In turn, this allows us to interpret how the kinematic algebra is made manifest in the case of
four-dimensional (anti-)self-dual solutions, thus resolving a long-standing conceptual issue in the
double copy literature. \\

The structure of our paper is as follows. In section~\ref{sec:power},
we detail our attempts to find power-like solutions of Euclidean
biadjoint scalar field theory in various numbers of spacetime
dimension, showing that non-trivial examples are absent in
$d=4$. In section~\ref{sec:EH}, we interpret this result as being due to
known properties of the Eguchi-Hanson instanton. In
section~\ref{sec:ansatz}, we present a more general ansatz for
double-copying instanton solutions. In section~\ref{sec:Cheung}, we
explain how our results relate to the recent two-dimensional
non-perturbative double copy proposal of ref.~\cite{Cheung:2022mix},
and clarify certain conceptual issues in the four-dimensional
non-perturbative double copy. Finally, we discuss our results and
conclude in section~\ref{sec:discuss}.

\section{Solutions of Euclidean biadjoint scalar theory}
\label{sec:power}

In Lorentzian $(-,+,+,+)$ signature, the equation of motion for
biadjoint scalar theory is as follows:
\begin{equation}
  \partial^2\Phi^{aa'}+y
  f^{abc}\tilde{f}^{a'b'c'}\Phi^{bb'}\Phi^{cc'}=0.
  \label{BAS}
\end{equation}
Here $\Phi^{aa'}$ is a scalar field with two colour indices in the
adjoint representation of two global gauge groups, whose structure
constants are $f^{abc}$ and $\tilde{f}^{a'b'c'}$
respectively. Furthermore, $y$ is a coupling constant. The quadratic
interaction term in eq.~(\ref{BAS}) arises from a cubic Lagrangian,
such that the energy of biadjoint scalar theory is unbounded from
below. This in turn makes it a not particularly physical theory, given
that its solutions will be dynamically unstable. However, this does
not prevent us from looking for such solutions, as in previous instances
of figure~\ref{fig:theories} for both scattering amplitudes and
classical solutions, quantities in biadjoint theory are indeed related
to physically well-behaved objects in gauge and gravity theory. Thus,
it is worthwhile and meaningful to classify solutions in biadjoint
theory, regardless of whether or not they make sense if considered in
isolation. \\

We may proceed to Euclidean signature in eq.~(\ref{BAS}) by
analytically continuing the time coordinate $t\rightarrow \tau=it$, such
that one has
\begin{equation}
    \partial^2\rightarrow \Delta,
    \label{Delta}
\end{equation}
where $\Delta$ is the Laplacian operator, and we work in $d$ spacetime
dimensions in general. We then have
\begin{equation}
  \Delta\Phi^{aa'}+y
  f^{abc}\tilde{f}^{a'b'c'}\Phi^{bb'}\Phi^{cc'}=0,
  \label{BAS2}
\end{equation}
for which we may attempt to find solutions as in the Lorentzian case
of ref.~\cite{White:2016jzc}. First, we take the gauge groups to
coincide, such that $f^{abc}=\tilde{f}^{abc}$. We may also write
\begin{equation}
  f^{abc}f^{a'bc}=T_A\delta^{aa'},
  \label{TAdef}
\end{equation}
where the constant $T_A$ depends on the common gauge group $G$, and
the normalisation of the generators. Next, we restrict to
spherically symmetric solutions via the ansatz
\begin{equation}
  \Phi^{aa'}=\frac{\delta^{aa'}}{y T_A} f(r),\quad
  r^2=x_{\mu}x^{\mu}.
  \label{Phiansatz}
\end{equation}
Substituting this into eq.~(\ref{BAS2}), one obtains
\begin{equation}
  \frac{1}{r^{d-1}}\frac{d}{dr}\left(r^{d-1}\frac{df(r)}{dr}\right)+f^2(r)=0,
\label{feq}
\end{equation}
where we have used the known form of the Laplacian in $d$-dimensional
spherical polar coordinates:
\begin{equation}
  \Delta f=\frac{1}{r^{d-1}}\frac{\partial}{\partial r}
  \left(r^{d-1}\frac{\partial f}{\partial r}\right)
  +\frac{1}{r^2}\Delta_{S^{d-1}}f.
  \label{Deld}
\end{equation}
The second term contains the {\it Laplace-Beltrami operator}
$\Delta_{S^{d-1}}$, which depends only upon angular coordinates. It
thus does not contribute to eq.~(\ref{feq}) given the dependence upon
only the radial coordinate. \\

Let us now look for pure power-like solutions by stipulating
\begin{equation}
  f(r)=A r^{\alpha},
  \label{powerlike}
\end{equation}
for some constants $A$ and $\alpha$. Substituting this into
eq.~(\ref{feq}) yields
\begin{equation}
  A\alpha(d+\alpha-2)r^{\alpha-2}+A^2 r^{2\alpha}=0.
  \label{powersol1}
\end{equation}
If this is to be true for all $r>0$, then we must have
\begin{equation}
  \alpha=-2\quad\Rightarrow\quad A[A-2(d-4)]=0.
  \label{alphaA}
\end{equation}
We thus find $A=0$ or $A=2(d-4)$, such that translating back to
eq.~(\ref{Phiansatz}) implies that there are two power-like solutions
in general. The first is not interesting -- it is the trivial (vacuum)
solution $\Phi^{aa'}=0$. The second is a non-trivial power-like
solution
\begin{equation}
  \Phi^{aa'}=\frac{2\delta^{aa'}}{y T_A}\frac{d-4}{r^2}.
  \label{Phisol2}
\end{equation}
Interestingly, the power of $r^{-2}$ is common to all spacetime
dimensions, which can be confirmed from dimensional analysis: the
dimensions of the field $\Phi^{aa'}$ and the coupling constant $y$
both vary with the number of dimensions, in just such a way as to fix
the power of radial distance for solutions involving an inverse power
of the coupling. This is in contrast to solutions of the {\it
  linearised} field equation, whose power of distance must vary in
order to maintain the correct dimensions of the field. A consistency
check of eq.~(\ref{Phisol2}) is that it reproduces the Lorentzian
monopole solutions of ref.~\cite{White:2016jzc} for $d=3$. These are
static solutions, and thus the field equation of eq.~(\ref{BAS})
reduces to that of eq.~(\ref{BAS2}), with a three-dimensional
Laplacian involving the spatial coordinates.\\

Arguably the most curious feature of eq.~(\ref{Phisol2}) is the
presence of $(d-4)$ in the numerator, which tell us that the
non-trivial power-like solution of the full Euclidean biadjoint scalar
field equation is absent in four spacetime dimensions. To gain more
insight into what is going on, it is instructive to examine more
general spherically symmetric solutions, as was done for the
Lorentzian case in ref.~\cite{DeSmet:2017rve}. That paper looked for
solutions in which the divergence of the biadjoint field at the origin
was (partially) screened.\footnote{Finite energy static solutions of
biadjoint scalar theory are impossible, as a consequence of {\it
  Derrick's theorem}~\cite{Derrick:1964ww}.} This motivates the form
\begin{equation}
  f(r)=\frac{K(r)-1}{r^2},
  \label{Kdef}
\end{equation}
which we are always entitled to write, and for which $K(r)\rightarrow
1$ (everywhere) constitutes the trivial solution. By further
introducing the variable $\xi$ via
\begin{equation}
  r=e^{-\xi},\quad -\infty <\xi < \infty,
\label{xidef}
\end{equation}
one may show that eq.~(\ref{feq}) amounts to
\begin{equation}
  \frac{\partial^2 K}{\partial\xi^2}-(d-6)\frac{\partial K}{\partial \xi}
  +(K-1)(K-2d+7)=0.
\label{Keq}
\end{equation}
This is a non-linear second-order differential equation, which cannot
be solved analytically in general.\footnote{By a further
transformation, one may recast eq.~(\ref{Keq}) as an {\it Abel
  equation of the second kind}, albeit not one that has a tractable
solution in terms of known functions.} However, we may visualise
solutions as follows. Defining $\psi\equiv \partial K/\partial\xi$, we
may write eq.~(\ref{Keq}) as two coupled first-order equations:
\begin{equation}
  \left(\frac{\partial K}{\partial\xi},\frac{\partial \psi}{\partial\xi}
  \right)=\Big(\psi,(d-6)\psi -(K-1)(K-2d+7)\Big).
  \label{Keq2}
\end{equation}
This defines a vector field in the $(K,\psi)$ plane, whose integral
curves correspond to solutions of eq.~(\ref{Keq}).
\begin{figure}
  \begin{center}
    \scalebox{0.5}{\includegraphics{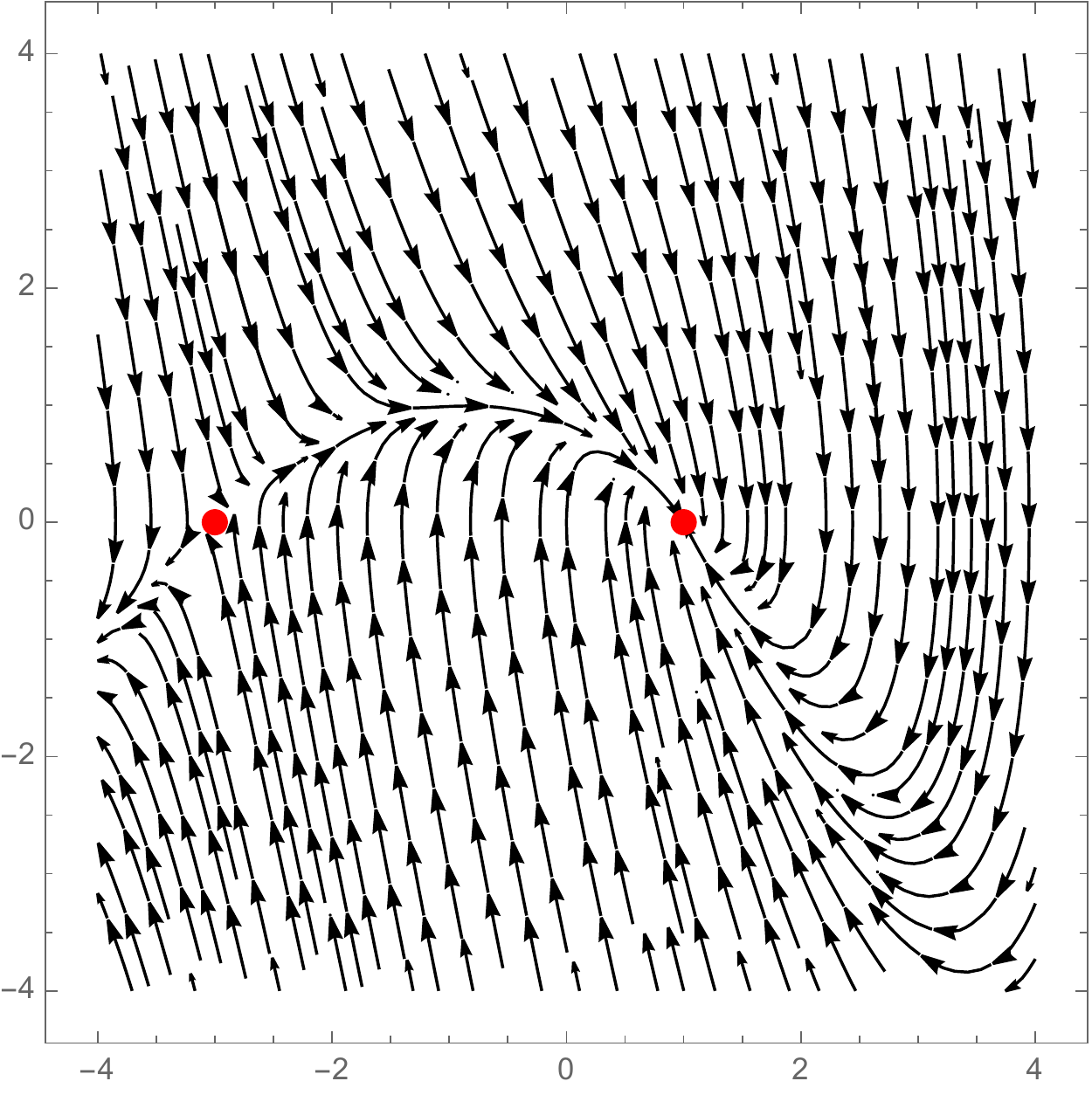}}
    \hspace{0.5cm}
    \scalebox{0.5}{\includegraphics{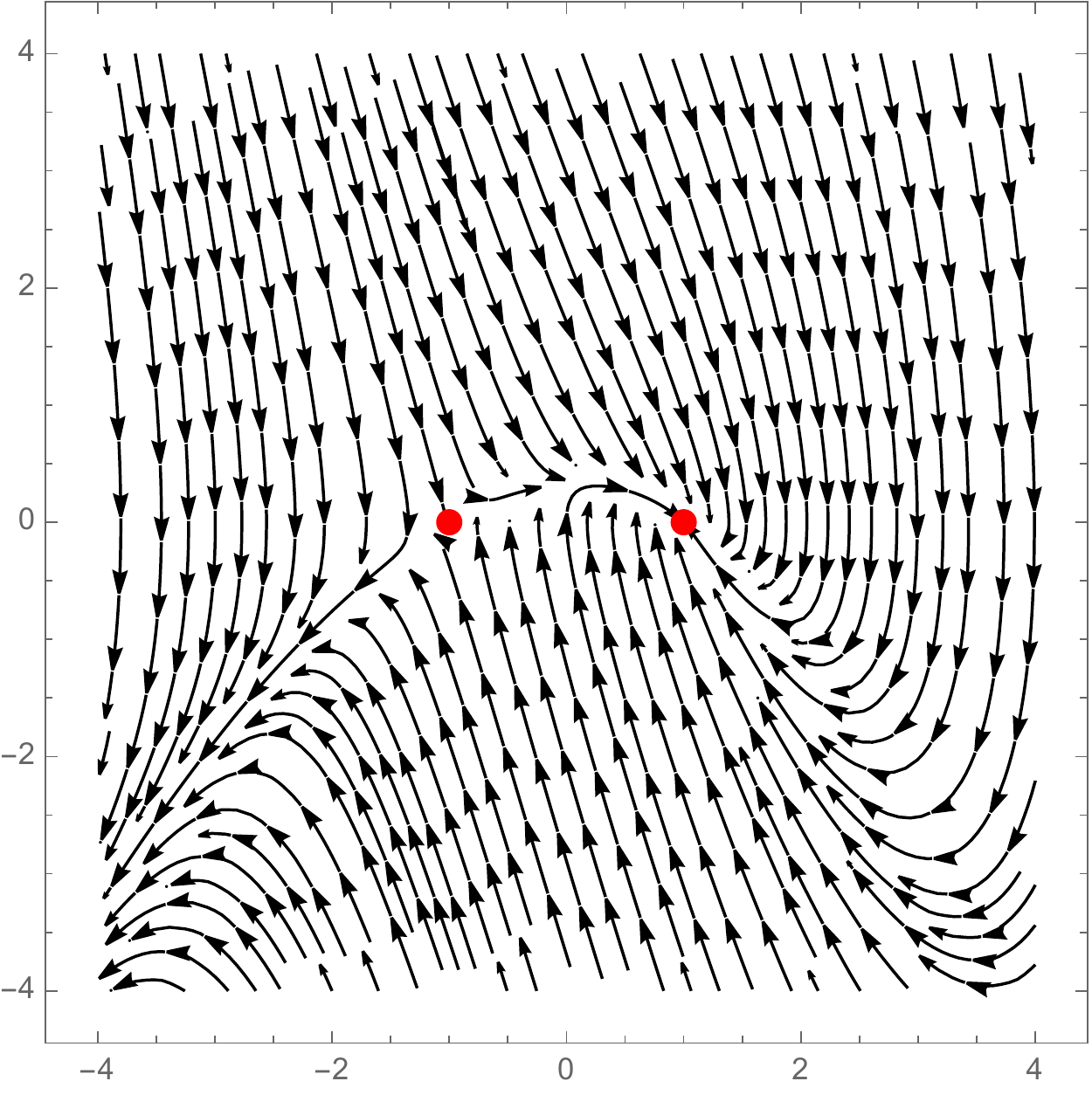}}\\
    \vspace{1cm}
    \scalebox{0.5}{\includegraphics{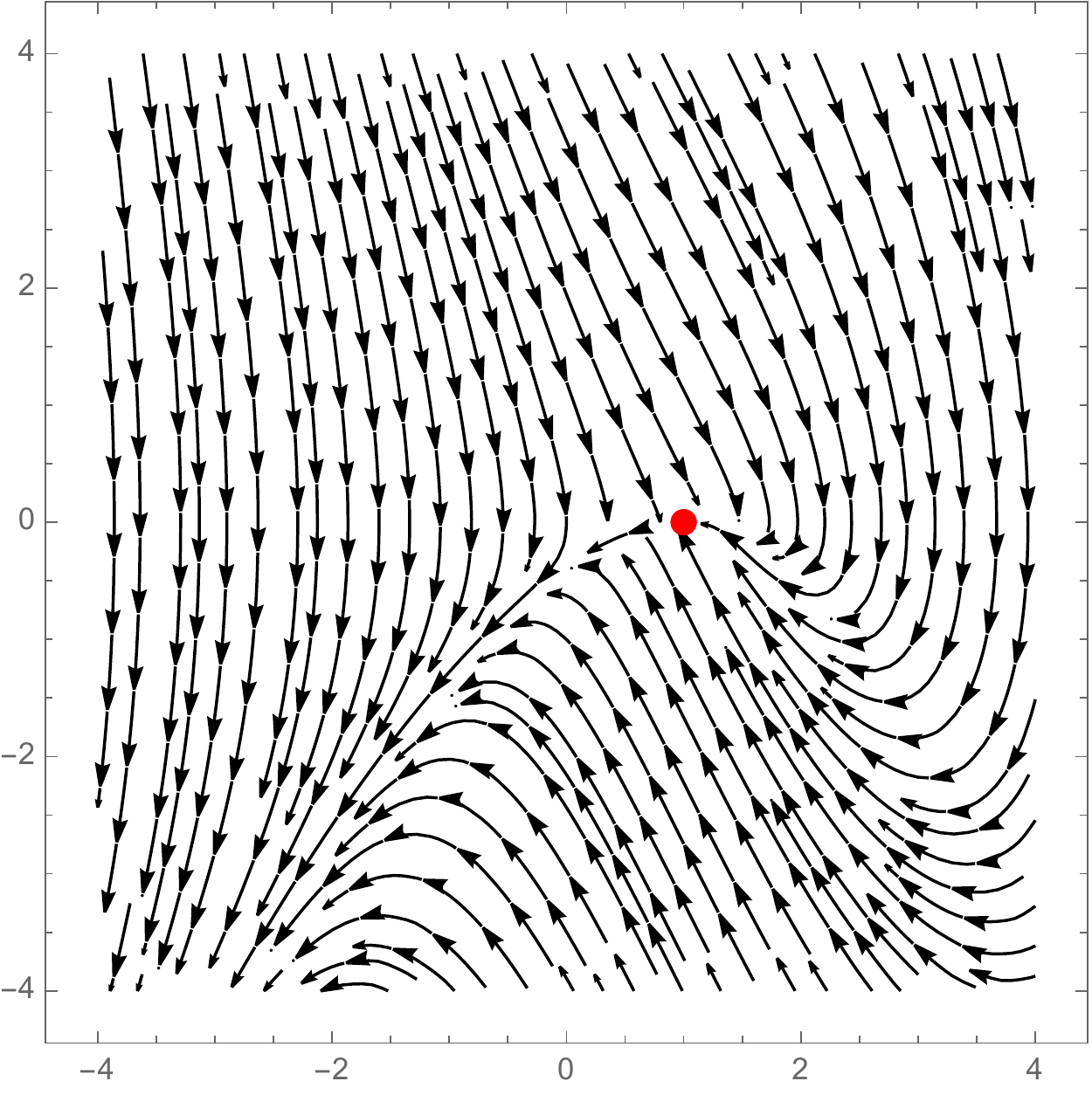}}
    \hspace{0.5cm}
    \scalebox{0.5}{\includegraphics{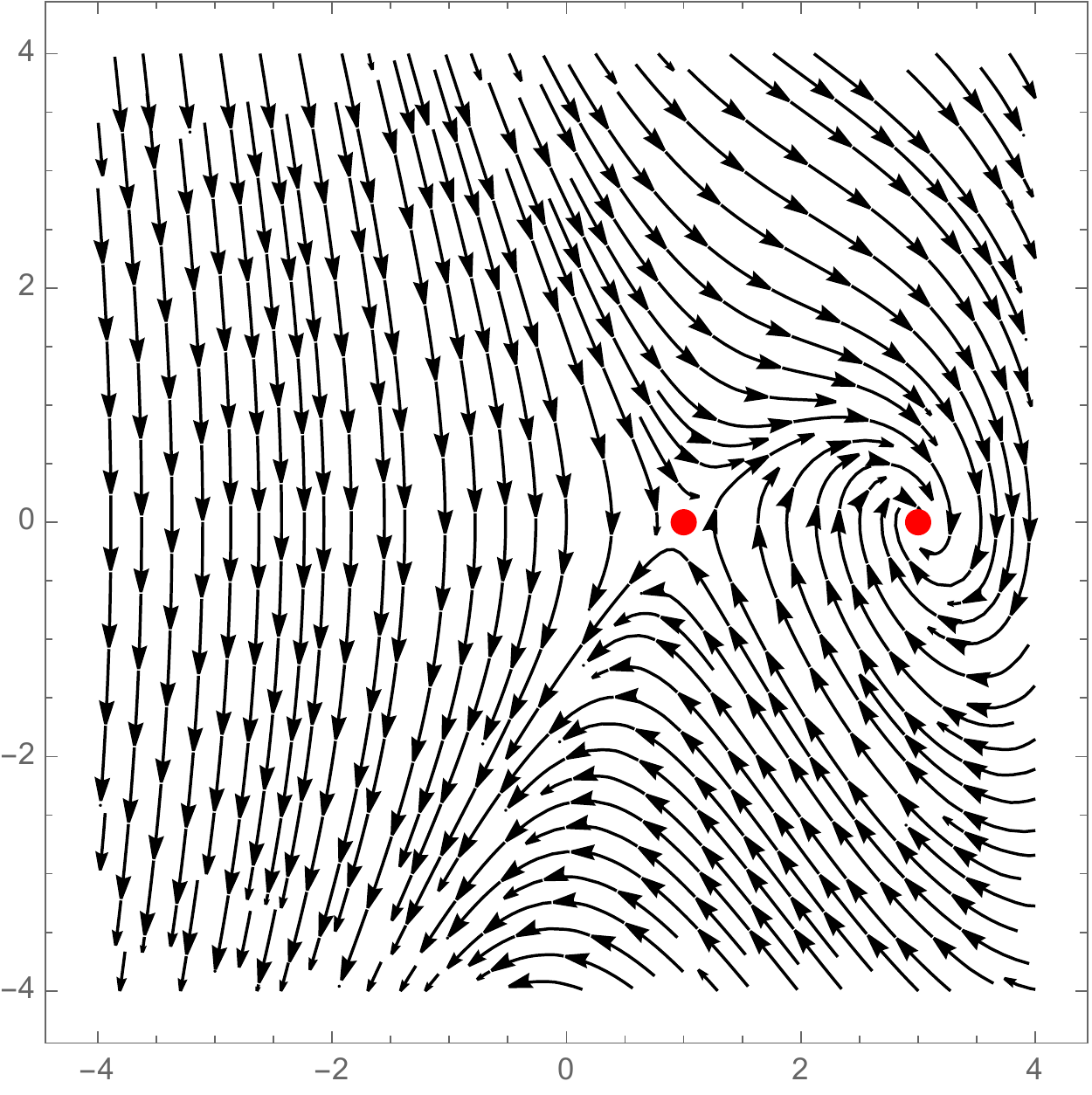}}
    \caption{Integral curves of the vector field of eq.~(\ref{Keq2})
      in the $(K,\psi)$ plane, corresponding to solutions of
      eq.~(\ref{Keq}). Shown are the cases $d=2$, 3, 4, and 5 respectively. The red dots correspond to the fixed
      point solutions $K=1$ and $K=2d-7$.}
    \label{fig:stream}
  \end{center}
\end{figure}
We show such curves for the cases of $d=2$, 3, 4 and 5 in
figure~\ref{fig:stream}. For general $d$, there are fixed points for
\begin{equation}
  K\in\{1,2d-7\},
  \label{Kvals}
\end{equation}
as is evident from eq.~(\ref{Keq}). Indeed, these correspond to the
trivial solution and the non-trivial power-like solution of
eq.~(\ref{Phisol2}) respectively. As the number of dimensions
increases from $d<4$, the non-trivial solution moves to the right in
the $(K,\psi)$ plane. For precisely $d=4$, the two fixed points
coincide, so that there is only the trivial solution, as found
above.\\

If we want to find solutions that partially screen the divergence at
the origin, we must look for bounded curves in the $(K,\psi)$ plane
i.e. those that correspond to finite numerators in
eq.~(\ref{Kdef}). For $d\neq 4$, there is always precisely one such
bounded curve, connecting the points $(1,0)$ and $(2d-7,0)$. Thus,
there is a single extended spherically symmetric solution that
corresponds to a screened charge. This broadens the implications of
the absence of a second fixed point in the $d=4$ case: not only is
there no non-trivial power-like solution, but there are are no
non-trivial extended solutions of the type in eq.~(\ref{Kdef}) either
(with bounded numerators). \\

In this section, we have undertaken a first investigation of the
spectrum of solutions of Euclidean biadjoint scalar field theory. We
find non-trivial power-like solutions in all spacetime dimensions
$d\neq 4$, with concomitant extended solutions. These solutions
deserve further study, but for the remainder of this paper we will
explain fully why there are no non-trivial power-like solutions for
$d=4$, and examine related implications.

\section{The Eguchi-Hanson instanton revisited}
\label{sec:EH}

In the previous section, we saw that there are no non-trivial
power-like solutions (or spherically symmetric solutions) of Euclidean
biadjoint theory in $d=4$. There is in fact a very simple reason why
this is the case. First, we may recall the observation made above that
non-linear power-like solutions always have a radial dependence $\sim
r^{-2}$, where this power can be fixed by dimensional analysis for
solutions that involve an inverse power of the coupling constant
$y$. However, in $d=4$, $r^{-2}$ is a harmonic function, whose
Laplacian vanishes for $r\neq 0$. To see this, note that in $d$
dimensions, eq.~(\ref{Deld}) implies
\begin{equation}
  \Delta r^{-n}= n(n+2-d)r^{-n-2},
  \label{Deltarn}
\end{equation}
which indeed vanishes if $r>0$ for $n=2$ and $d=4$ (at $r=0$, there is
a singularity, leading to an appropriately normalised delta function
on the right-hand side of eq.~(\ref{Deltarn})). If $r^{-2}$ is
harmonic in $d=4$, this means that it solves the {\it linearised}
biadjoint equation of eq.~(\ref{BAS2}). There is thus no room left for
it to solve the non-linear equation, which is why there is no
non-trivial power-like solution of Euclidean biadjoint theory in four
spacetime dimensions. Indeed, we could simply have started with this
observation, and not bothered with the analysis of the previous
section at all. We maintain, however, that the results of the previous
section remain useful: there are indeed non-trivial power-like
solutions in other numbers of dimensions. Furthermore, the observation
that there are no bounded extended solutions of Euclidean biadjoint
theory, as well as power-like forms, is itself interesting.\\

As is well-known~\cite{Monteiro:2014cda}, one can turn harmonic
functions into solutions of the full biadjoint scalar theory by
dressing them with constant colour vectors $\{c^a,\tilde{c}^{a'}\}$,
which in the case of our power-like solution in $d=4$ becomes
\begin{equation}
  \Phi^{aa'}=\frac{\alpha c^a \tilde{c}^{a'}}{r^2},
  \label{EHzeroth}
\end{equation}
where we have included an arbitrary constant $\alpha$ that remains
unfixed by the requirement that the kinematic dependence is
harmonic. It is straightforward to check that, upon substitution into
eq.~(\ref{BAS2}), the non-linear term vanishes, leaving only the linear
term as required. Given that previous examples of the classical double
copy have focused on solutions that linearise biadjoint theory, we can
then ask if it is possible to identify the gauge and gravity solutions
for which eq.~(\ref{EHzeroth}) constitutes the zeroth copy. Indeed,
the solution turns out to be already known: it is related to the {\it
  Eguchi-Hanson (EH) solution} in gravity. First derived and discussed
in refs.~\cite{Eguchi:1978xp,Eguchi:1978gw,Eguchi:1979yx}, this is a
solution whose finite energy, self-dual nature and asymptotically
Euclidean character lead to its interpretation as a {\it gravitational
  instanton}. As pointed out in
e.g. refs.~\cite{Tod,Berman:2018hwd,Luna:2018dpt}, it is particularly
convenient to express the EH solution in (2,2) signature, using the
coordinate system
\begin{equation}
  u=\frac{\tau-iz}{\sqrt{2}},\quad
    v=\frac{\tau+iz}{\sqrt{2}},\quad X=\frac{ix-y}{\sqrt{2}},\quad
    Y=\frac{ix+y}{\sqrt{2}},
    \label{uvXY}
\end{equation}
in terms of Euclidean Cartesian coordinates\footnote{As is common in the instanton literature, we let Greek indices take the values $\mu,\nu,...= 1,2,3,4$, with $x_4=\tau$.}
\begin{equation}
x_\mu=(x,y,z,\tau).   
  \label{Cartesian}
\end{equation}
Then the EH metric may be written as
\begin{equation}
g_{\mu\nu}=\eta_{\mu\nu}+h_{\mu\nu},
\label{hdef}
\end{equation}
where $h_{\mu\nu}$ is the graviton field
\begin{equation}
  h_{\mu\nu}=\phi k_\mu k_\nu,\quad \phi=\frac{\lambda}
  {(uv-XY)}, \quad k_\mu=\frac{1}{(uv-XY)}(v,0,0,-X).
  \label{EHsol}
\end{equation}
The metric of eqs.~(\ref{hdef}, \ref{EHsol}) is in so-called {\it
  Kerr-Schild form}, where the vector $k_\mu$ satisfies the null and
geodesic conditions
\begin{equation}
k^2=0\quad k\cdot \partial k_\mu=0.
\label{KSconditions}
\end{equation}
This in turn means that its single and zeroth copies may be
straightforwardly taken, using the general procedure defined in
ref.~\cite{Monteiro:2014cda}. One simply writes
\begin{equation}
  A_\mu^a=c^a \phi k_\mu,\quad \phi^{aa'}=c^a\tilde{c}^{a'}\phi,
  \label{singlezeroth}
\end{equation}
where $c^a$ and $\tilde{c}^a$ are constant colour vectors. Then, the
gauge and biadjoint fields thus constructed are guaranteed to solve
the Yang-Mills and biadjoint equations, which happen to linearise in
both cases. Upon translating to (Euclidean) Cartesian coordinates in
eq.~(\ref{EHsol}), one finds
\begin{equation}
  \phi=\frac{2\lambda}{r^2},
\end{equation}
so that the zeroth copy of the Eguchi-Hanson solution in
eq.~(\ref{singlezeroth}) precisely matches the power-like solution of
eq.~(\ref{EHzeroth}) as claimed, provided we
identify\footnote{Alternatively, one can absorb the arbitrary constant
$\alpha$ into the additional colour vector $\tilde{c}^{a'}$ that
appears in the biadjoint field.} $\alpha=2\lambda$. \\

There is another way to interpret the above results. First writing the
gauge field of eq.~(\ref{singlezeroth}) in terms of an abelian gauge
field $A_\mu$:
\begin{equation}
  A_\mu^a=c^a A_\mu,\quad A_\mu=\frac{\lambda}{(uv-XY)^2}
  \left(v,0,0,-X\right),
  \label{AmuEH}
\end{equation}
we may recognise the latter as
\begin{equation}
  A_\mu=\hat{k}_\mu \phi,\quad \hat{k}_\mu=-(\partial_u,0,0,\partial_Y),
  \label{khatdef}
\end{equation}
such that the single copy is given by the action of a differential
operator on $\phi$. It is easily checked that the Eguchi-Hanson
graviton is given in terms of the same operator:
\begin{equation}
  h_{\mu\nu}=\hat{k}_\mu \hat{k}_\nu \phi,
\label{EHdiff}
\end{equation}
such that the double copy is formulated as a product in momentum, rather than
position, space. A similar idea has occurred in the literature
before. Reference~\cite{Monteiro:2014cda} pointed out that setting up
the double copy in terms of differential operators reproduces known
descriptions of (anti-)self-dual Yang-Mills theory and gravity. That
is, for a $\hat{k}_\mu$ satisfying the two conditions
\begin{equation}
  \hat{k}^2=0,\quad \partial\cdot\hat{k}=0,
  \label{khatconditions}
\end{equation}
substituting the ansatz of eq.~(\ref{EHdiff}) into the Einstein
equations reduces the latter to the {\it Plebanski equation} of
self-dual gravity. For a suitable choice of $\hat{k}_\mu$, this can be
written in the lightcone coordinate system as
\begin{equation}
  \partial^2\phi+\kappa\left\{\partial_Y\phi,\partial_u\phi\right\}=0,
  \label{Plebanski1}
\end{equation}
where the Poisson bracket is defined by
\begin{equation}
  \{A,B\}=(\partial_Y A)(\partial_u B)-(\partial_u A)(\partial_Y B).
  \label{Poisson}
\end{equation}
Alternatively, one may write the Plebanski equation directly in terms
of a general operator $\hat{k}_\mu$ satisfying
eq.~(\ref{khatconditions}),
as~\cite{Monteiro:2014cda}\footnote{Following convention, we do not
raise or lower indices given that we are in Euclidean signature.}
\begin{equation}
  \partial^2\phi-\frac{1}{2}(\hat{k}_\mu\hat{k}_\nu\phi)
  (\partial_\mu\partial_\nu\phi)=0.
\label{Plebanski2}
\end{equation}
Similarly, substituting the ansatz
\begin{equation}
  A_\mu^a=\hat{k}_\mu \Phi^a,
  \label{gaugetheorykhat}
\end{equation}
into the Yang-Mills equations, for $\hat{k}_\mu$ satisfying
eq.~(\ref{khatconditions}), leads to a known formulation of self-dual
Yang-Mills theory~\cite{Parkes:1992rz}\footnote{Our eq.~(\ref{SDYM})
can be obtained from eq.~(28) of ref.~\cite{Monteiro:2014cda}. Our
conventions differ in that we are in Euclidean signature. We have also
set coupling constants to unity in both eqs.~\eqref{Plebanski2} and~\eqref{SDYM}, and normalised the vector
$\hat{k}_\mu$ such that numerical constants are the same in all
theories, for reasons that will become clear later on.}
\begin{equation}
  \partial^2 \Phi^a-\frac{1}{2}\epsilon^{abc}
  (\hat{k}_\mu\Phi^b)(\partial_\mu\Phi^c)=0,  
\label{SDYM}
\end{equation}
where the anti-self-dual sectors of both gauge and gravity theory can
be obtained similarly (i.e. by a different choice of
$\hat{k}_\mu$). The (anti-)self dual sectors constitute explicit
cases in which the double copy can be made manifest at the level of
equations of motion~\cite{Monteiro:2011pc}, and generalisations of this construction to
exotic deformed theories are also known~\cite{Chacon:2020fmr}. The
Eguchi-Hanson instanton is a special case in which the fields are of
Kerr-Schild form. In the gauge theory, this specialises the ansatz of
eq.~(\ref{gaugetheorykhat}) to
\begin{equation}
  \Phi^a=c^a \phi,
  \label{Phiaphi}
\end{equation}
which linearises the Yang-Mills equations, such that one may consider
the abelian field $A_\mu$ of eq.~(\ref{AmuEH}). As noted in
ref.~\cite{Berman:2018hwd}, the observation that (anti-)self-dual
gravity solutions can be defined in terms of differential operators,
and associated with (null) electromagnetic fields, was made long ago
in ref.~\cite{Tod}. The double copy reinterprets this observation, and
provides a framework for potential generalisations.\\

Returning to the present study, it is not obvious that the double copy
between eqs.~(\ref{AmuEH}, \ref{khatdef}) and eq.~(\ref{EHdiff}) is a special case of the
known self-dual double copy of ref.~\cite{Luna:2018dpt}, as the $\hat{k}_\mu$ operator of
eq.~(\ref{khatdef}) does not satisfy both conditions in
eq.~(\ref{khatconditions}). For our purposes, it will be useful to transform the
operator to Cartesian coordinates, in which it takes the form\footnote{In
an abuse of notation, we will refer to the operator as $\hat{k}_\mu$
in both the $(x,y,z,\tau)$ and $(u,v,X,Y)$ coordinate systems, given that
the explicit coordinates that appear in any given equation imply no
ambiguity.}
\begin{equation}
  \hat{k}_\mu=-\frac{1}{2}\left(\partial_x+i\partial_y,
  \partial_y-i\partial_x,\partial_z-i\partial_{\tau},
  \partial_{\tau}+i\partial_z\right),
  \label{khatCartesian}
\end{equation}
which may be written more compactly as
\begin{equation}
  \hat{k}_\mu=-\frac{1}{2}\left(\delta_{\mu\nu}+i\bar{\eta}^3_{\mu\nu}\right)
  \partial_\nu.
  \label{khat2}
\end{equation}
Here $\bar{\eta}^3_{\mu\nu}$ is a special case of the {\it 't Hooft
  symbols} $\{\bar{\eta}^a_{\mu\nu}\}$, which arise in the study of
instantons. For convenience, we review the properties of these
symbols, as well as useful identities, in
appendix~\ref{app:thooft}. More briefly, the 't Hooft symbols
$\{\bar{\eta}^a_{\mu\nu}\}$ form a representation of an SU(2)
subalgebra of SO(4), where the latter group is
equivalent to the Lorentz group in Euclidean signature. Thus, the
presence of the 't Hooft symbol in eq.~(\ref{khat2}) means that the
operator $\hat{k}_\mu$ involves a particular 
``rotation'' of the derivative operator $\partial_\nu$. Using
eq.~(\ref{etaids}), it is then straightforward to verify that
\begin{equation}
  \hat{k}^2=0,\quad \partial\cdot \hat{k}=-\frac12 \Delta,
\end{equation}
and thus that the second condition in eq.~(\ref{khatconditions}) is
not satisfied. However, one may instead consider the alternative
differential operator
\begin{equation}\label{kPrimeLC}
 \hat{k}'_{\mu} = (0,\partial_Y,\partial_u,0),
\end{equation}
In Cartesian coordinates, this translates as
\begin{equation}\label{kOpPrime}
  \hat{k}'_{\mu} 
  = \frac{1}{2}\left(\bar{\eta}^2_{\mu\nu} - i\bar{\eta}^1_{\mu\nu}\right)\partial_{\nu}
  =-\bar{\eta}^2_{\mu\nu}\hat{k}_{\nu},
\end{equation}
which does indeed satisfy both of the properties in
eq.~(\ref{khatconditions}). Furthermore, we see that $\hat{k}'_{\mu}$
is a coordinate transformation of $k_\mu$ to a new frame whose
coordinates are
\begin{equation}
  {x'}_\mu=-\bar{\eta}^2_{\mu\nu}x_\nu\quad\Rightarrow\quad
  \left(\begin{array}{c}x'\\y'\\z'\\\tau'
  \end{array}\right)=
    \left(\begin{array}{c}z\\\tau\\-x\\-y
  \end{array}\right).
\end{equation}
We then have
\begin{equation}
  (r')^2={x'}_\mu\,{x'}_\mu={x}_\mu\,{x}_\mu=r^2,
\end{equation}
such that one may write the Eguchi-Hanson solution and its single copy
in the primed coordinate system as
\begin{equation}
  A'_\mu=c^a \hat{k}'_\mu \phi(r'),\quad h'_{\mu\nu}=\hat{k}'_\mu
  \hat{k}'_\nu \phi(r'),
\label{A'h'}
\end{equation}
which is indeed a special case of the general self-dual construction
of eqs.~(\ref{EHdiff}--\ref{gaugetheorykhat}). \\

In this section, we have explained the absence of non-linear
power-like solutions in Euclidean biadjoint scalar theory, by pointing
out that simple power-like solutions in four dimensions are in fact
already ``claimed'' by the zeroth copy of the Eguchi-Hanson instanton,
for which the biadjoint field equations linearise. A natural language
for describing the single and zeroth copies of the EH solution is in
terms of differential operators satisfying eq.~(\ref{khatconditions}),
such that one obtains a special case of the self-dual double copy
construction proposed in ref.~\cite{Monteiro:2014cda}. Particularly
compelling in our present paper is the fact that differential
operators satisfying eqs.~(\ref{khatconditions}) can be written in
terms of 't Hooft symbols, as in eq.~(\ref{kOpPrime}). This suggests a
general ansatz for double-copying certain instanton solutions, that we
explore in the following section. We will also find that one may
easily construct {\it non-abelian} single copies of the Eguchi-Hanson
instanton, thus making its single copy structure more intricate than
previously thought.

\newpage
\section{A general ansatz for double-copying instantons}
\label{sec:ansatz}

In the previous section, we discussed the single copy of the
Eguchi-Hanson instanton, which can be taken to be an abelian-like
gauge field. In general, however, there are many instanton solutions
of non-abelian gauge theories, namely (anti-)self-dual classical
solutions, of finite energy (see
e.g. refs.~\cite{Weinberg:2012pjx,Manton:2004tk} for pedagogical
reviews). To be concrete, let us consider the case of SU(2) gauge
theory. Finite energy demands that the field become pure gauge at
infinity, and a given solution then constitutes a map from the
boundary of spacetime ($S^3$) to the gauge group manifold, which is
also $S^3$ for SU(2). Instantons can then be classified by their {\it
  winding}, or {\it instanton number}, which has a simple
interpretation as the number of times the first $S^3$ space wraps
around the second, in mapping the two
manifolds.\footnote{Mathematically, one talks about the {\it third
homotopy group} of a manifold ${\cal M}$, which classifies
non-trivial maps from $S^3$ to a given manifold ${\cal M}$. Choosing
${\cal M}$ to be the gauge group manifold for SU(2), one has
$\pi_3(S^3)={\mathbb Z}$, meaning that there are topologically
distinct maps labelled by different integers. This is precisely the
winding number mentioned above, where positive (negative) values
correspond to the (anti-)self dual sectors.} The winding number is given by the volume integral
\begin{equation}
k=\frac{1}{16\pi^2}\int d^4x {\rm Tr}\left[\tilde{\bf F}_{\mu\nu}\,
{\bf F}_{\mu\nu}\right],
\label{knum}
\end{equation}
where the field strength ${\bf F}_{\mu\nu}\equiv F_{\mu\nu}^a{\bf T}^a$ and its dual $\tilde{\bf F}_{\mu\nu}\equiv \tilde{F}_{\mu\nu}^a{\bf T}^a$ are given respectively by
\begin{equation}
  F_{\mu\nu}^a = \partial_{\mu}A_{\nu}^a - \partial_{\nu}A_{\mu}^a + \epsilon^{abc}A_{\mu}^bA_{\nu}^c, \qquad \tilde{F}^a_{\mu\nu} = \frac{1}{2}\epsilon_{\mu\nu\rho\sigma}F^a_{\rho\sigma}.
\end{equation}
There will be a non-trivial parameter space of solutions for definite
winding number $k$ in general, with the various parameters
representing the width or size of the solution, its position in space,
rotation angles in spacetime or in the gauge space etc. Some of these
parameters are redundant under gauge transformations or other
redundancies, but the set of independent parameters that label
instantons of given $k$ are called {\it moduli}, and they form a {\it
  moduli space}. The metric in this space is not completely smooth,
but can be singular at certain points e.g. for instantons that have
zero size, or where the centres of multiple instantons
coincide. Remarkably, it is known how to classify {\it all} possible
instanton solutions in pure YM theory~\cite{Atiyah:1978ri}. \\

\begin{figure}
  \begin{center}
    \scalebox{0.6}{\includegraphics{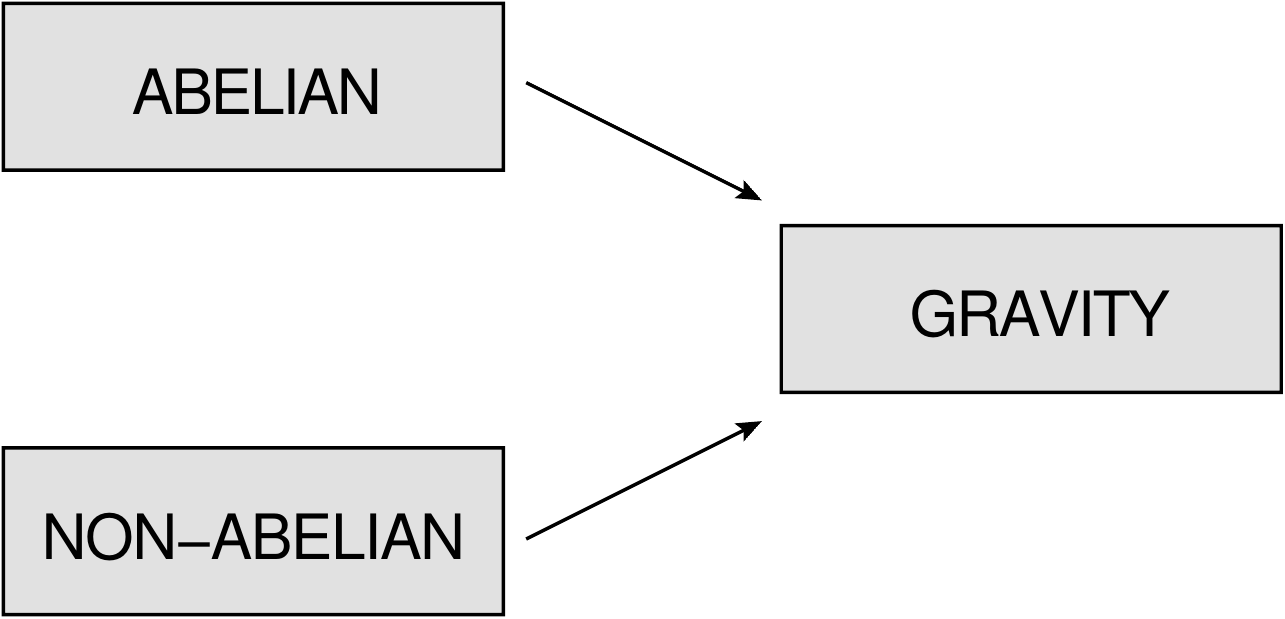}}
    \caption{Schematic depiction of the multiple single copies of the
      gravitational instanton solutions considered in this paper. The
      same scheme has previously been obtained for magnetic
      monopoles~\cite{Bahjat-Abbas:2020cyb,Alfonsi:2020lub}, and for
      infrared singularities of scattering
      amplitudes~\cite{Oxburgh:2012zr}.}
    \label{fig:abelnonabel}
  \end{center}
\end{figure}

A large family of SU(2) instanton solutions is given by dressing a
vector field $V_\mu$ according to the {\it 't Hooft
  ansatz}~\cite{tHooft:1976snw}
\begin{equation}
A_\mu^a=-\bar{\eta}^a_{\mu\nu} V_\nu,\quad 
A_\mu^a=-\eta^a_{\mu\nu} V_\nu,
\label{thooft}
\end{equation}
for the case of self-dual and anti-self dual fields
respectively, and where
$\{\eta_{\mu\nu}^a,\bar{\eta}_{\mu\nu}^a\}$ are the 't Hooft symbols
encountered above, and reviewed briefly here in
appendix~\ref{app:thooft}.\footnote{The minus signs in eq.~(\ref{thooft}) are
conventional. Also, note that use of the anti-self-dual matrix
$\bar{\eta}^a_{\mu\nu}$ actually results in a self-dual field. Whether
or not $\eta^a_{\mu\nu}$ or $\bar{\eta}^a_{\mu\nu}$ appears in the
gauge field can depend upon the gauge.} In our previous use of a 't Hooft symbol in
eq.~(\ref{khat2}), this was acting merely as a representation of a
particular spacetime rotation, where the upper index labelled which
particular infinitesimal rotation we were talking about. In
eq.~(\ref{thooft}), however, the upper indices on the 't Hooft symbols
are to be interpreted as adjoint indices associated with the SU(2)
gauge group. That this is possible is due to the fact that the
$\{\eta^a_{\mu\nu}\}$ and $\{\bar{\eta}^a_{\mu\nu}\}$ separately form
  SU(2) algebras of SO(4) rotations. They can thus be mapped to the
  SU(2) gauge algebra. Focusing on the self-dual case, substitution of
  eq.~(\ref{thooft}) into the Yang-Mills equations yields the
  condition
\begin{equation}
\partial_\mu V_\mu+V_\mu V_\mu=0,
\label{Vmucond1}
\end{equation}
as well as 
\begin{equation}
\tilde{f}_{\mu\nu}=f_{\mu\nu},\quad f_{\mu\nu}=\partial_\mu V_\nu-
\partial_\nu V_\mu.
\label{Vmucond2}
\end{equation}
Here we can recognise $f_{\mu\nu}$ as being analogous to an
abelian-like field strength tensor. However, $V_\mu$ cannot
necessarily be interpreted as an abelian gauge field, given that
$f_{\mu\nu}$ is not guaranteed to satisfy the Maxwell equation
\begin{equation}
  \partial_\mu f_{\mu\nu}=0
  \label{Maxwell}
\end{equation}
in general. Equation~(\ref{Vmucond1}) is usually satisfied by taking
$V_{\mu}$ to be the gradient of the logarithm of a harmonic function:
\begin{equation}
V_\mu=\partial_\mu\log V,\quad \Delta V=0.
\label{VlogV}
\end{equation}
However, motivated by the discussion in the previous section, there is
another ansatz we can make. Let us construct a vector field according to the prescription:
\begin{equation}
  A_\mu=\hat{k}_\mu \phi,\quad \hat{k}_\mu=\left(A\delta_{\mu\nu}+B_i\,
    \bar{\eta}^i_{\mu\nu}\right)\partial_\nu,\quad i\in \{1,2,3\},
  \label{Vmuansatz}
\end{equation}
where the scalar $A$ and three-vector $B_i$ are possibly complex constants. This generalises the definitions of $\hat{k}_\mu$ and
$\hat{k}'_\mu$ given by eqs.~(\ref{khat2}, \ref{kOpPrime}). Note that,
as in eq.~(\ref{khat2}), the upper index on each 't Hooft symbol is
not a gauge (adjoint) index, but merely labels which infinitesimal
rotations we are talking about. With this field we can compute
\begin{equation}
  f_{\mu\nu}=B_i\left(\bar{\eta}^i_{\nu\rho}\partial_\mu
    -\bar{\eta}^i_{\mu\rho}\partial_\nu\right)\partial_\rho\phi,
  \label{fmunu}
\end{equation}
from which one finds
\begin{equation}
  \tilde{f}_{\mu\nu}=f_{\mu\nu}+B_i\bar{\eta}^i_{\mu\nu}\Delta\phi.
  \label{ftildeexp}
\end{equation}
Thus, we see that the requirement for $A_{\mu}$ to satisfy the self-dual
equations in eq.~\eqref{Vmucond2} is such as to force the field $\phi$ to be a harmonic
function ($\Delta\phi=0$). If we further insist that the double copy
of eq.~(\ref{Vmuansatz}), defined according to eq.~(\ref{EHdiff}), be
a solution of self-dual gravity, then the differential operator
$\hat{k}_\mu$ must satisfy the dual constraints of
eq.~(\ref{khatconditions}). For the first, we find
\begin{equation}
  \hat{k}^2=(A^2+B^2)\Delta=0\quad\Rightarrow\quad
  A^2=-B^2,
  \label{k2ansatz}
\end{equation}
where $B^2 = B_iB^i$ and we have used the identities in appendix~\ref{app:thooft}. The
second condition gives
\begin{equation}
  \partial_\mu \hat{k}_\mu=A\Delta=0.
  \label{A=0}
\end{equation}
Thus, the requirement that the abelian self-dual gauge field of
eq.~(\ref{Vmuansatz}) double-copies to a self-dual gravity
solution imposes
\begin{equation}
  A=0,\quad B^2=0.
  \label{ABvals}
\end{equation}
The components $\{B_i\}$ are then required to be complex in general.
Indeed, the abelian single copy of Eguchi-Hanson, with a differential
operator defined as in eq.~(\ref{kOpPrime}), emerges as a special
case. Interestingly, though, eq.~(\ref{ABvals}) implies that
eq.~(\ref{Vmucond1}) is automatically satisfied: one finds
\begin{equation}
  \partial_\mu A_\mu+A_\mu A_\mu=A\Delta\phi+(A^2+B^2)
  (\partial_\mu\phi)(\partial_\mu\phi)=0.
\end{equation}
This immediately implies that we may dress the solution of
eq.~(\ref{Vmuansatz}) to form a non-abelian SU(2) instanton, as in
eq.~(\ref{thooft}): 
\begin{align}
  A_\mu^a&=-\bar{\eta}^a_{\mu\nu}\hat{k}_\nu\phi.
  \label{Vmuansatz2}
\end{align}
A suggestive way to write this is as
\begin{equation}
  A_\mu^a=-\hat{k}^a_\mu\phi,\quad
  \hat{k}^a_\mu\equiv\bar{\eta}^a_{\mu\nu}\hat{k}_\nu,
  \label{kahat}
\end{equation}
i.e. in terms of a ``non-abelian'' differential
operator. Intriguingly, the non-abelian instanton can be double-copied
directly, upon tracing over the colour indices: using
eq.~(\ref{etaids}), one finds
\begin{equation}
  \hat{k}_\mu^a \hat{k}_\nu^a=-\hat{k}_\mu\hat{k}_\nu.
  \label{kmuknu}
\end{equation}
Hence, if we construct the graviton field
\begin{equation}
  h_{\mu\nu}=-\hat{k}^a_{\mu}\hat{k}^a_\nu\phi,
  \label{graviton2}
\end{equation}
this yields precisely the same gravity solution as double copying the
field of eq.~(\ref{Vmuansatz}) according to eq.~(\ref{EHdiff}). As
noted above, eq.~(\ref{Vmuansatz}) is not guaranteed to be an abelian gauge field
in general, as it may not necessarily satisfy the Maxwell equation of
eq.~(\ref{Maxwell}). In fact it does, however, due to the consequence
derived above that $\phi$ be harmonic. To see this, note that
eq.~(\ref{fmunu}) implies
\begin{equation}
  \partial_\mu f_{\mu\nu}=B_i\bar{\eta}^i_{\nu\rho}\partial_\rho (\Delta \phi)
  =0,
\end{equation}
where we have used the antisymmetry property
$\bar{\eta}^i_{\mu\nu}=-\bar{\eta}^i_{\nu\mu}$. We therefore find that
one may construct either an {\it abelian} or {\it non-abelian} single
copy of certain (anti-)self-dual gravity solutions, for which the
Eguchi-Hanson solution discussed in the previous section is a special
case. Pleasingly, this mirrors the situation that has been found for
magnetic monopole solutions in
refs.~\cite{Bahjat-Abbas:2020cyb,Alfonsi:2020lub}. That is, one may
regard the single copy of the pure NUT solution in gravity as an
abelian-like (Dirac) magnetic monopole, dressed by a constant colour
vector, or as a genuinely non-abelian Wu-Yang monopole, where there is
a singular gauge transformation that relates the two forms. This
scheme is shown in figure~\ref{fig:abelnonabel}, and is interesting to
note that a similar idea has occurred before in the original double
copy for scattering amplitudes. For example, the infrared
singularities of both abelian and non-abelian gauge theory map to the
{\it same} infrared singularities in gravity, to all orders in
perturbation theory~\cite{Oxburgh:2012zr}. There is presumably a gauge
transformation (that we have not been able to find) that relates the
two forms of the single copy instantons considered here.\\

In this section, we have given a general ansatz for single-copying a class of gravitational instantons, that allows us to construct non-abelian as
well as abelian single copies. However, all of the solutions thus
obtained turn out to be special, in that they linearise the equations
of motion in both gravity and gauge theory (and hence, by association,
biadjoint theory). The question then arises of whether one can
single-copy more general instantons, or more formally: what portion of
the moduli space of gauge theory instantons is captured by the ansatz
of eq.~(\ref{Vmuansatz2})? The answer to the latter question appears
to be rather limited. For example, we may attempt to calculate the
winding number integral of eq.~(\ref{knum}), for the single copy of
the Eguchi-Hanson solution. This turns out to yield\\
\begin{equation}
  {\rm Tr}[\tilde{\bf F}_{\mu\nu}{\bf F}_{\mu\nu}] \propto
  \frac{\lambda^2B^2}{r^8},
\end{equation}
which vanishes due to being proportional to $B^2=0$. Thus, the
non-abelian single copy of the Eguchi-Hanson instanton is
topologically trivial.\footnote{Similar conclusions were reached for
the abelian single copy in ref.~\cite{Berman:2018hwd}, albeit using a
different single copy field. See ref.~\cite{Luna:2018dpt} for a
discussion of the relation between the single copy of
ref.~\cite{Berman:2018hwd}, and that used in this paper.} Also, as a
singular solution whose singularity appears non-removable by a gauge
transformation, it should not properly be regarded as part of the
moduli space of gauge theory instantons. Similar conclusions will be
reached for more general functions $\phi$: the winding number cannot
depend on which particular basis of the rotation generators entering
$\hat{k}_\mu$ we choose i.e. the 't Hooft symbols
$\{\bar{\eta}^i_{\mu\nu}\}$. This in turn implies invariance under
rotations of the vector $B_i$, such that the result can only
depend on $B^2$, which is trivial. Interestingly, though, the
space of harmonic functions $\phi$ in gauge theory still allows for
some potentially interesting gravity solutions. For example, one may
choose
\begin{equation}
  \phi=\sum_{i=1}^N \frac{c_i}{(x-a_i)^2},
  \label{phisol2}
\end{equation}
which has $N$ point-like disturbances at 4-positions $\{a_i\}$. The
double copy of this would appear to be a multi-centre generalisation of
the Eguchi-Hanson solution. Given that the single-centre Eguchi-Hanson
case is related to a two-centre Gibbons-Hawking
metric~\cite{Gibbons:1979xm,Prasad:1979kg}, the graviton obtained from
eqs.~(\ref{EHdiff}, \ref{phisol2}) may also be of multi-centre
Gibbons-Hawking type. We have not been able to find an explicit
coordinate transformation that realises this, which by no means rules
out that such a transformation is possible. \\

Of course, we have not considered the most general ansatz for gauge
theory solutions in this paper. Equation~(\ref{Vmuansatz2}) relies on
a single function $\phi$ which is dressed by additional factors,
rather than the full adjoint-valued field $\phi^a$ of
eq.~(\ref{gaugetheorykhat}). However, it is not clear how to generate
the additional information required by eq.~(\ref{gaugetheorykhat})
upon taking the single copy, i.e. how to turn the single gravitational
function $\phi$ into the multiple
functions $\{\phi^a\}$.\footnote{Reference~\cite{Campiglia:2021srh} recently
considered the more general ansatz of eq.~(\ref{gaugetheorykhat}) in
exploring how asymptotic symmetries in self-dual Yang-Mills and
gravity are related by the double copy. However, replacements between
theories were at the level of commutators, or equivalently structure
constants.} A prescription for achieving something
similar has recently been proposed in various two-dimensional
theories~\cite{Cheung:2022mix}. Indeed, comparing our analysis in this
paper with the results of ref.~\cite{Cheung:2022mix} yields a number
of useful insights, which we explore in the following section.

\section{Relation to the two-dimensional non-perturbative double copy}
\label{sec:Cheung}

In the previous section, we provided an ansatz for single-copying
exact solutions of (anti-)self-dual gravity. Recently, another
non-perturbative single copy procedure has
appeared~\cite{Cheung:2022mix}, and the aim of this section is to
compare their approach with our analysis in this paper. As we will
see, this comparison reveals a number of useful insights, that clarify
long-standing questions in the double copy literature, but also
generalise the results of ref.~\cite{Cheung:2022mix} itself. The
latter reference considered various field theories in two spacetime
dimensions, including the biadjoint scalar field theory we have
already encountered in this paper. In the conventions of
ref.~\cite{Cheung:2022mix}, and allowing for arbitrary gauge groups,
this has equation of motion
\begin{equation}
  \partial^2\phi^{aa'}-\frac12 f^{abc}\tilde{f}^{a'b'c'}\phi^{bb'}\phi^{cc'}
  =0,
  \label{BASCheung}
\end{equation}
in Lorentzian signature. Solutions in this theory were argued to obey
a similar scheme to figure~\ref{fig:theories}, but where the gauge
theory is replaced by {\it Zakharov-Mikhailov (ZM)}
theory~\cite{Zakharov:1973pp}:
\begin{equation}
  \partial^2 \phi^a-\frac12 f^{abc}\partial_\mu\phi^a\tilde{\partial}^\mu
  \phi^b=0.
  \label{ZM}
\end{equation}
Here the field $\phi^a$ carries a single adjoint index, and we have
introduced the dual derivative operator
\begin{equation}
  \tilde{\partial}^\mu=\epsilon^{\mu\nu}\partial_\nu,
  \label{dualpartial}
\end{equation}
where $\epsilon^{\mu\nu}$ is the two-dimensional Levi-Civita
symbol. Finally, the gravity theory is replaced by {\it Special
  Galileon (SG)
  theory}~\cite{Cheung:2014dqa,Cheung:2015ota,Hinterbichler:2015pqa},
whose equation of motion is
\begin{equation}
  \partial^2\phi-\frac12(\partial_\mu\partial_\nu\phi)
  (\tilde{\partial}^\mu\tilde{\partial}^\nu\phi)=0.
  \label{SG}
\end{equation}
There is a clear pattern of replacements in proceeding from
eq.~(\ref{BASCheung}) through eqs.~(\ref{ZM}, \ref{SG}): adjoint
indices are progressively removed, and the number of spacetime (and
dual) derivatives increases. More formally, given two adjoint-valued
fields $V^a$ and $W^a$, one may define the formal replacement rules
\begin{equation}
  V^a\rightarrow V,\quad f^{abc} V^a W^b\rightarrow (\partial_\mu V)
  (\tilde{\partial}^\mu W),
\label{VWreplace}
\end{equation}
which can indeed be used to transform between the various theories
mentioned above. The second replacement can be interpreted as
replacing the structure constants of the colour algebra with those of
an infinitely dimensional kinematic algebra. The kinematic structure
constants are more easily viewed in momentum space, and to find them
we may Fourier transform the right-hand side of the second replacement
in eq.~(\ref{VWreplace}) to get
\begin{align}
  \int d^2 x e^{ip_1\cdot x} \partial_\mu V\tilde{\partial}^\mu W&=
  \int d^2 x \int\frac{d^2 p_2}{(2\pi)^2}\int\frac{d^2 p_3}{(2\pi)^2}
  e^{i(p_1-p_2-p_3)\cdot x} [-\epsilon^{\mu\nu}p_{2\mu}p_{3\nu}]
  \tilde{V}(p_2)\tilde{W}(p_3) \nonumber\\
  &=\int\frac{d^2 p_2}{(2\pi)^2}\int\frac{d^2 p_3}{(2\pi)^2}
        [-\epsilon^{\mu\nu}p_{2\mu}p_{3\nu}\delta^2(p_2+p_3-p_1)]\tilde{V}(p_2)
        \tilde{W}(p_3),
\end{align}
where we have introduced momentum modes of the spacetime fields,
denoted with tildes. We thus see that eq.~(\ref{VWreplace}) replaces
adjoint-valued fields with scalars, which are then combined according
to the momentum-space kinematic structure constant
\begin{equation}
  {f_{p_2p_3}}^{p_1}=X(p_2,p_3)\delta^2(p_2+p_3-p_1),\quad
  X(p_2,p_3)=-\epsilon^{\mu\nu}p_{2\mu}p_{3\nu}.
\label{kinstruc}
\end{equation}
As to the nature of this kinematic algebra, it describes
area-preserving diffeomorphisms of the two-dimensional spacetime. To
see this, we may write the general form of an such an infinitesimal
diffeomorphism in two-dimensional spacetime:
\begin{equation}
  {\bf V}=-(\tilde{\partial}^\mu V)\partial_\mu.
  \label{Vdef}
\end{equation}
According to standard differential geometry results, an infinitesimal
diffeomorphism $f^\mu \partial_\mu$ is area-preserving (in more than
two dimensions, volume-preserving) provided that $\partial_\mu
f^\mu=0$. Equation~(\ref{dualpartial}) implies
\begin{equation}
\partial_\mu \tilde{\partial}^\mu V=0,
\end{equation}
so that the diffeomorphism of eq.~(\ref{Vdef}) is area-preserving as
claimed. As explained in ref.~\cite{Cheung:2022mix}, evaluating the
Lie algebra of the generators in eq.~(\ref{Vdef}) yields
\begin{equation}
  [{\bf V},{\bf W}]={\bf Z},\quad {\bf Z}=-\tilde{\partial}^\mu
  \left(\partial_\nu V\tilde{\partial}^\nu W\right)\partial_\mu.
\end{equation}
To recognise the above structure constants, we can take generators
corresponding to individual momentum modes:
\begin{equation}
  {\bf V}_{p_i}=-\tilde{\partial}^\mu (V_{p_i})\partial_\mu,\quad
  V_{p_i}=e^{ip_i\cdot x},
\end{equation}
from which one finds
\begin{equation}
  [{\bf V}_{p_2},{\bf V}_{p_3}]=X(p_2,p_3){\bf V}_{p_2+p_3}
  ={f_{p_2p_3}}^{p_1}{\bf V}_{p_1},
\end{equation}
where we have recognised the form of the structure constant of
eq.~(\ref{kinstruc}). \\

Above, we have seen that one may transform between the theories of
eqs.~(\ref{BASCheung}, \ref{ZM}, \ref{SG}) by successively replacing
colour algebras with a Lie algebra of area-preserving
diffeomorphisms. This does not explain {\it why} one should make such
replacements, however, and ref.~\cite{Cheung:2022mix} provided the
following motivation. It is a known fact that the group U($N$) becomes
isomorphic, at large $N$, to the group of diffeomorphisms of a
torus~\cite{Hoppe:1988gk}. To make this precise, one may use the fact
that there is a particular basis $\{{\bf T}_p\}$ of the generators of
U($N$) (for odd $N$), where $p$ is a 2-vector whose components are
integer modulo $N$, and such that the structure constants
are~\cite{Hoppe:1988gk,Cheung:2022mix}
\begin{equation}
  {f_{p_2 p_3}}^{p_1}=-\frac{N}{2\pi}\sin\left(\frac{2\pi}{N}\epsilon^{\mu\nu}
  p_{2\mu}p_{3\nu}\right).
  \label{U(N)}
\end{equation}
Upon taking the large $N$ limit, one may reinterpret the vectors
$\{p_i\}$ as specifying momentum modes on a torus, and the structure
constants of eq.~(\ref{U(N)}) reproduce precisely those of
eq.~(\ref{kinstruc}). Thus, in the large $N$ limit, the three theories
of eqs.~(\ref{BASCheung}, \ref{ZM}, \ref{SG}) (for gauge group U($N$),
repeated in the case of biadjoint theory) become mutually
isomorphic. Reference~\cite{Cheung:2022mix} then uses this to argue
that the colour-kinematic replacements of eq.~(\ref{VWreplace}) should
be made also for finite $N$, and also proposes a scheme for turning
non-perturbative solutions of SG theory into counterparts in ZM and
biadjoint theory, which is accurate up to subleading corrections in
$N$.\\

\begin{figure}
\begin{center}
\scalebox{0.6}{\includegraphics{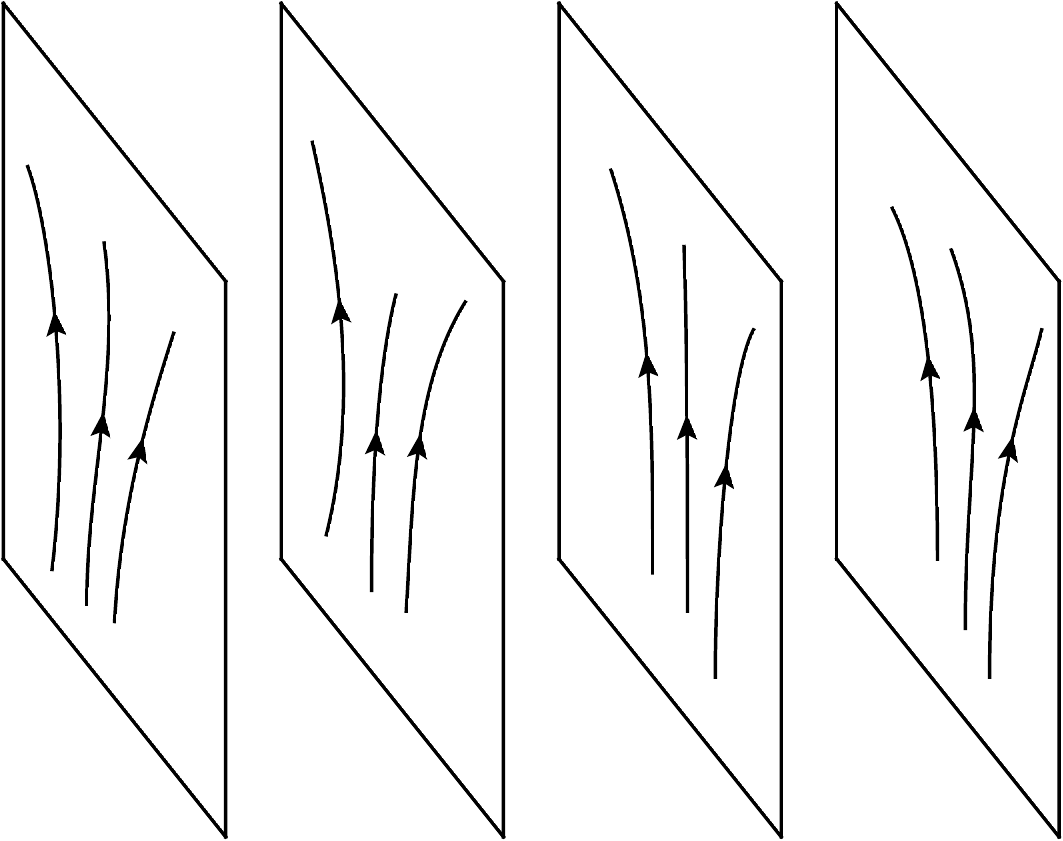}}
\caption{Foliation of four-dimensional Euclidean space by a family of
  $\alpha$- or $\beta$-planes. The abelian single copy gauge field
  $A_\mu$ generates area-preserving diffeomorphisms in each plane,
  represented here by field lines (integral curves).}
\label{fig:toastrack}
\end{center}
\end{figure}

Regardless of this motivation, a very similar scheme -- albeit perhaps
not noticeably so -- has appeared in the literature before, namely in
the study of (anti-)self dual Yang-Mills and
gravity~\cite{Monteiro:2011pc}. We have quoted the relevant field
equations in eqs.~(\ref{SDYM}, \ref{Plebanski2}), where different
choices of the differential operator $\hat{k}_\mu$ correspond to the
(anti-)self-dual cases respectively. Comparing these with
eqs.~(\ref{ZM}, \ref{SG}), we see that
four-dimensional (anti-)self-dual Yang-Mills theory and gravity have
{\it precisely} the same forms as two-dimensional ZM and SG theory
respectively, but where the dual derivative operator
$\tilde{\partial}^\mu$ is replaced by the differential operator
$\hat{k}^\mu$.\footnote{It should now hopefully be clear
why we have chosen different conventions in our
eqs.~(\ref{Plebanski2}, \ref{SDYM}) relative to those in existing
literature~\cite{Monteiro:2014cda}: it is to make the similarity
between our four-dimensional theories and the two-dimensional
theories of ref.~\cite{Cheung:2022mix} more striking.} Indeed, as is the role of $\tilde{\partial}^\mu$ in two
dimensions, we can associate $\hat{k}_\mu$ with area-preserving
diffeomorphisms, where the most general such transformation will be given by
\begin{equation}
  {\bf V}=-(\hat{k}^\mu V)\partial_\mu.
  \label{Vkhat}
\end{equation}
That this is area-preserving follows from our imposition that
$\partial\cdot \hat{k}=0$, and the cases of self-dual and
anti-self-dual YM theory or gravity arise, as explained above, from
the general ans\"{a}tze
\begin{equation}
  \hat{k}_\mu\Big|_{\rm SD}=\frac{1}{2}B_i \bar{\eta}^i_{\mu\nu}\partial^\nu,\quad
  \hat{k}_\mu\Big|_{\rm ASD}=\frac{1}{2}B_i \eta^i_{\mu\nu}\partial^\nu,\quad
\end{equation}
respectively. The role of the vector $B_i$ is to pick out the
two-dimensional planes in which the area-preserving diffeomorphisms of
eq.~(\ref{Vkhat}) act. Specifically, we may write
\begin{equation}
  B_i\bar{\eta}^i_{\mu\nu}=b^{(1)}_{[\mu}b^{(2)}_{\nu]},
  \label{bivec1}
\end{equation}
where the explicit forms of the vectors on the right-hand side, as may
be verified using eq.~(\ref{thooftmatrices}), are
\begin{equation}
  b^{(1)}_\mu=\left(B_1,B_2,B_3,0\right)\quad
  b^{(2)}_\mu=\left(0,\frac{B_3}{B_1},\frac{-B_2}{B_1},-1\right),
  \label{b12}
\end{equation}
where we have assumed $B_1\neq 0$ and $B^2 = 0$ as before. Then 
\begin{equation}
  (\hat{k}^\mu\phi)\partial_\mu
  =(b^{(1)[\mu} b^{(2)\nu]}\partial_\nu\phi)\partial_\mu
  \label{bivec2}
\end{equation}
will generate diffeomorphisms in the plane defined by the {\it
  bivector} $b^{(1)}_{[\mu}b^{(2)}_{\nu]}$. As an example, we may
consider the self-dual operator of eq.~(\ref{kOpPrime}), which has
\begin{equation}
  B_1=-i,\quad B_2=1,\quad B_3=0.
\end{equation}
This gives (in Cartesian coordinates)
\begin{equation}
  b^{(1)}_{\mu}=(-i,1,0,0),\quad b^{(2)}_{\mu}=(0,0,-i,-1).
\end{equation}
Translating to the lightcone coordinate system, one finds
diffeomorphisms in the $(u,Y)$ plane as expected. Note that the
vectors $\{b^{(i)}\}$ satisfy the conditions
\begin{equation}
  b^{(i)}\cdot b^{(j)}=0,\quad\forall i\in\{1,2\}.
  \label{bconditions}
\end{equation}
This makes the plane defined by the above bivector an example of a
{\it null plane}. Null planes defined by self-dual and anti-self-dual
bivectors are called $\alpha$-planes and $\beta$-planes respectively,
and they play a key role in the study of instantons (see
e.g. ref.~\cite{nash1988topology} for a pedagogical review). To
summarise, we have found that (anti-)self-dual YM theory and gravity
provide a four-dimensional generalisation of the non-perturbative
double copy construction of ref.~\cite{Cheung:2022mix}, but where the
area-preserving diffeomorphisms take place in $\alpha$- or
$\beta$-planes. To visualise this over the entire space, we may
foliate four-dimensional Euclidean space by a family of parallel
$\alpha$- or $\beta$-planes, whose orientation is determined by the
vector $B_i$. The vector field $A_\mu$ will then generate
area-preserving diffeomorphisms in each one, as shown in
figure~\ref{fig:toastrack}.\\ 

It was already known that the kinematic
algebra of self-dual YM and gravity consisted of area-preserving
diffeomorphisms~\cite{Monteiro:2011pc}. New to this paper, however,
are the geometric construction of area-preserving diffeomorphisms in
arbitrary null planes, and the recognition that this provides a
four-dimensional analogue of the two-dimensional non-perturbative
double copy proposed in ref.~\cite{Cheung:2022mix}. Regarding the
latter, it would be possible, for example, to argue for the double
copy replacements
\begin{equation}
  \phi^a\rightarrow \phi,\quad f^{abc}\phi_1^b\phi_2^c\rightarrow
  \partial_\mu \phi_1\,\hat{k}^\mu\phi_2
  \label{khatreplace}
\end{equation}
on similar grounds to eq.~(\ref{VWreplace}). That is, one could take
the large $N$ limit of self-dual U($N$) Yang-Mills theory (or
U($N$)$\times$U($N$) biadjoint theory), consider periodic solutions in
the planes associated with $\hat{k}^\mu$, and then argue that the
theories become mutually isomorphic.\\

\begin{figure}
  \begin{center}
    \scalebox{0.6}{\includegraphics{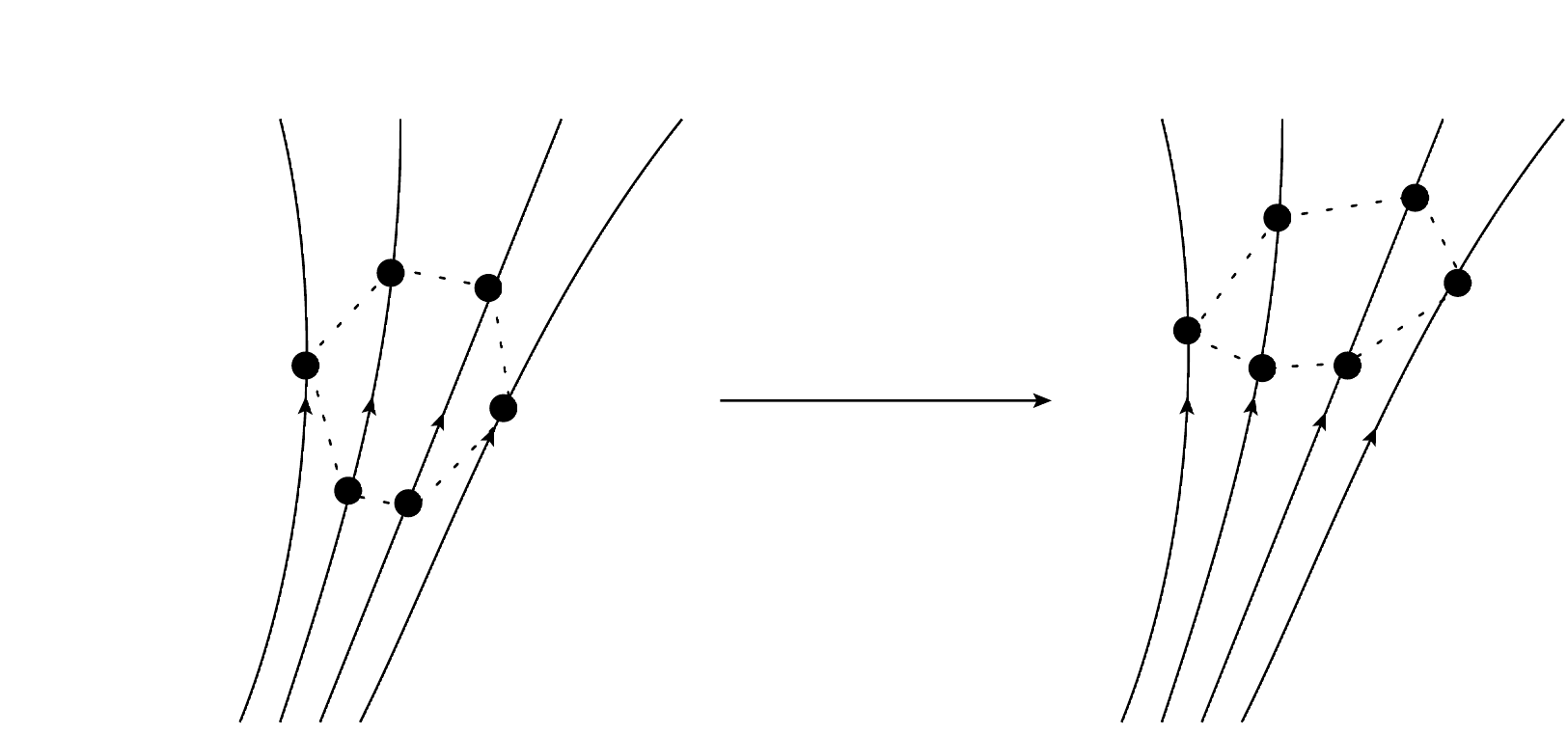}}
    \caption{A vector field defines a diffeomorphism consisting of
      simultaneous translations along the integral curves of the
      field. For an area-preserving diffeomorphism, the areas of the
      two shapes shown on the left and right will be the same.}
    \label{fig:diffeomorphism}
  \end{center}
\end{figure}

Although the presence of an area-preserving diffeomorphism algebra in
(anti-)self-dual gauge and gravity theory has been known for some
time, it has not been known how these transformations are visible or
relevant when considering exact classical solutions. For scattering
amplitudes, the situation is much clearer, as first explained in
ref.~\cite{Monteiro:2011pc}: amplitudes in all theories defined by
eqs.~(\ref{Plebanski2}, \ref{SDYM}, \ref{BASCheung}) are given by an
expansion in cubic diagrams, each of whose vertices involves a product
of two structure constants appropriate to the theory of
interest. Thus, amplitudes in one theory can simply be obtained from
amplitudes in another by replacing the appropriate structure
constants, which gives a direct operational meaning to phrases such as
``replacing the colour algebra with a kinematic algebra''. For
classical solutions, no structure constants manifestly appear, and
thus it is not clear how moving from one theory to another
involves a replacement of algebras. This conceptual problem is
especially pronounced given that BCJ duality for amplitudes is an
intrinsically non-linear phenomenon: it crucially relies on higher
orders in perturbation theory, such that products of structure
constants appear. The above ideas, however, indeed allow us to
interpret what is happening, even though our exact classical solutions
{\it linearise} the equations of motion. For a given gravity solution
of the form of eq.~(\ref{EHdiff}), let us choose its abelian single
copy
\begin{equation}
  A_\mu^a=c^a A_\mu,\quad A_\mu=\hat{k}_\mu\phi.
  \label{Amuabelian}
\end{equation}
Using standard results from differential geometry (see
e.g.~\cite{schutz_1980}), we may regard the vector field $A_\mu$ as
generating an infinitesimal diffeomorphism
\begin{equation}
  A^\mu\partial_\mu,
\end{equation}
whose physical interpretation is that it performs a simultaneous
translation along the integral curves of the field
(figure~\ref{fig:diffeomorphism}). Vector fields that are the single
copies of gravitational solutions will then generate diffeomorphisms
of form
\begin{equation}
  (\hat{k}^\mu\phi)\partial_\mu,
\end{equation}
which, as remarked above, generate area-preserving diffeomorphisms in
each of the null planes associated with the operator
$\hat{k}_\mu$. Thus, even for the case of linearised solutions that
do not involve higher-order contractions of structure constants, there
is still a well-defined way in which their properties are governed by
the area-preserving diffeomorphism algebra. Furthermore, translating
between biadjoint, gauge and gravity theory entails replacing the
diffeomorphism generators with colour generators, or vice versa. This
is straightforward to see in the case of solutions with abelian-like
single copies, which give rise to the following fields in different
theories (contracted with appropriate generators):
\begin{equation}
  \Phi=(c^a{\bf T}^a)\otimes(\tilde{c}^{a'}{\bf \tilde{T}}^{a'})\phi,
  \quad {\bf A}^{\mu} \partial_\mu=
  (c^a{\bf T}^a) (\hat{k}^\mu\phi)\partial_\mu,\quad h^{\mu\nu}\partial_\mu\partial_\nu
  =(\hat{k}^\mu\hat{k}^\nu\phi)\partial_\mu\partial_\nu.
\end{equation}
Upon proceeding from left to right, we can see directly that colour
generators are replaced by generators of area-preserving
diffeomorphisms. There is nothing particularly profound going on here:
the replacements of generators simply constitute the statement that
the field in each theory must be Lie-algebra valued in two Lie
algebras. These will be colourful or kinematical, as dictated by which
theory we are in. Notably, this correspondence works for any value of
$N$, not just the large $N$ limit. But, as explained above, it applies
only to those solutions that can be chosen to linearise their
respective field equations. Interestingly, the non-perturbative double
copy discussed in ref.~\cite{Cheung:2022mix} suffers from a similar
specialism, in that SG theory is ultimately related to a free field
theory.\\

Given the ideas of this section, it is not clear how to fully
interpret the non-abelian single copy of eq.~(\ref{Vmuansatz2}), which
does not obviously translate to replacing an area-preserving
diffeomorphism generator with a colour counterpart. It would be
interesting to find non-trivial examples of double copies of the form
of eq.~(\ref{gaugetheorykhat}), which in turn relates to the question
of whether all gauge theory instantons can be double-copied. It may
well turn out that, whilst all solutions of the Plebanksi equation can
be single-copied to make non-abelian instantons, the converse is not
possible. Indeed, similar comments were made in
ref.~\cite{Cheung:2022mix} regarding how all solutions of SG theory
can be used to make solutions of ZM theory, but that the opposite is
not true. Complementary statements were made, some time ago and
completely independently of the double copy, in
ref.~\cite{Oh:2011nv}. The authors showed that every gravitational
instanton solution can be mapped to a SU(2) non-abelian instanton,
albeit one that lives in the {\it same} Ricci-flat background as that
defined by the gravitational solution i.e. the SU(2) solutions are
``self-gravitating''. In double copy lingo, this corresponds to a
so-called {\it type B} curved-space double
copy~\cite{Bahjat-Abbas:2017htu}, in which a classical solution in
gravity can be identified with a gauge field living on a non-dynamical
curved background. Although this is a different situation to that
considered in this paper, it nevertheless has the property that not
all gauge theory solutions can be mapped to those in gravity. The
reason in this case, though, is unique to the particular set-up
considered in ref.~\cite{Oh:2011nv}: one cannot copy a SU(2) solution
on a particular curved background to obtain a gravity solution
corresponding to a {\it different} curved space.

\section{Discussion}
\label{sec:discuss}

In this paper, we have performed a first investigation of the spectrum
of non-linear solutions of biadjoint scalar field theory in four
Euclidean dimensions. Our motivation stems from the known double copy
relationships between various field theories, summarised here in
figure~\ref{fig:theories}. It is not known how general this scheme is,
and finding a genuinely non-perturbative incarnation would be a big
step forward in this regard. We found that the spectrum of Euclidean
solutions is rather rich in dimensions other than four, mirroring
similar results that have been obtained previously in Lorentzian
signature~\cite{White:2016jzc,DeSmet:2017rve,Bahjat-Abbas:2018vgo}. In
precisely four spacetime dimensions, however, there are no simple
power-like solutions, with a consequent absence of dressed solutions
that screen a power-like divergence at the origin. This can be traced
to the fact that the power-like form that is required is a harmonic
function in $d=4$, and thus solves the linearised biadjoint field
equation. It can be identified with the zeroth copy of the
Eguchi-Hanson solution, which has previously been considered from a
double copy point of view in
refs.~\cite{Berman:2018hwd,Luna:2018dpt}. The single and zeroth copies
can be formulated in terms of certain differential operators, obeying
conditions that are similar (but not the same) as the Kerr-Schild
conditions underlying the exact classical double copy of
ref.~\cite{Monteiro:2014cda}. \\

We have here reinterpreted these differential operators as a special
case of a more general ansatz, that involves the well-known 't Hooft
symbols that arise in the study of SU(2) gauge theory instantons. We
provide a geometric interpretation of this ansatz, showing that our
general (anti-)self dual operators are associated with area-preserving
diffeomorphisms in given families of null planes. Further use of the
't Hooft symbols allows us to construct exact non-abelian single
copies of (anti-)self-dual gravity solutions, living in an SU(2) gauge
theory. However, the gauge theory requirements of (anti-)self-duality
restrict the class of solutions to those that are ultimately
linearisable. Nevertheless, the presence of both abelian and
non-abelian single copies realises the same scheme
(figure~\ref{fig:abelnonabel}) that has previously arisen in the study
of magnetic monopoles~\cite{Bahjat-Abbas:2020cyb,Alfonsi:2020lub}.  \\

Our results make contact with a recently proposed non-perturbative
double copy in two spacetime dimensions. In particular, the
replacement of colour algebras by area-preserving diffeomorphism
algebras in four dimensions is a direct analogue of similar
replacements that relate biadjoint scalar, Zakharov-Mikhailov and
Special Galileon theory in two dimensions. Whilst the presence of such
algebras in the (anti-)self dual sectors of YM theory and gravity have
been known for some time~\cite{Monteiro:2011pc}, we show for the first
time how to interpret the replacement of colour by kinematic algebras
for exact classical solutions, rather than amplitudes. \\

Given that the full set of instanton solutions in non-abelian gauge
theories is known~\cite{Atiyah:1978ri}, the question arises of how
general our methods for explicitly double-copying them are. In more
formal language, it remains unclear how the moduli space of gauge
theory instantons maps onto that of gravitational
instantons. Intriguing in this regard is that the explicit examples of
solutions that we have given here seem to be excluded from the moduli
space of SU(2) instanton solutions. Based on similar conclusions in
other contexts~\cite{Cheung:2022mix,Oh:2011nv}, it seems likely that
not all gauge theory instantons can be double-copied to make
gravitational solutions. However, it is correct to say that a full
classification of {\it which} solutions can be copied is both useful,
and missing. \\

There are many potential avenues for further work. One might look for
non-spherically symmetric solutions of Euclidean biadjoint theory, and
also interpret the spectrum of existing solutions we have found in
various numbers of dimension, including their relationship with the
Lorentzian solutions of
refs.~\cite{White:2016jzc,DeSmet:2017rve,Bahjat-Abbas:2018vgo}. Examining
how general our methods are -- in terms of mapping out the known
moduli space of gauge theory instantons -- would be useful, and a
first step in this regard would perhaps be to try to make sense of the
more general ansatz of eq.~(\ref{gaugetheorykhat}), as noted
explicitly in ref.~\cite{Campiglia:2021srh}. Finally, we note that
twistor methods are ubiquitous in the study of instantons, and have
recently arisen in the context of the exact classical double
copy~\cite{White:2020sfn,Farnsworth:2021wvs,Chacon:2021wbr,Chacon:2021lox,Chacon:2021hfe,Guevara:2021yud,Adamo:2021dfg}. Some
sort of twistorial description of the (anti-)self-dual double copy is
surely possible. We look forward to reporting on these various topics
in the future.


\appendix

\section{'t Hooft symbols and their properties}
\label{app:thooft}

In this appendix, we introduce and briefly review the properties of
{\it 't Hooft symbols}, which are used in the study of instanton
solutions in non-abelian gauge theories. They are used in two
different ways throughout this paper. First, they arise as
infinitesimal generators of Euclidean rotations (the Euclidean
signature equivalents of Lorentz transformations), which constitute
the group SO(4). This group is six-dimensional, comprising three
rotations $\{J_i\}$ in the $(x_i,x_j)$ plane, and three rotations
$\{K_i\}$ in the $(x_i,x_4)$ plane, where $i\in\{1,2,3\}$, and we have
defined coordinates as in eq.~(\ref{Cartesian}). Note that the latter are
the analogue of boosts in Lorentzian signature. One may then form the
combinations
\begin{equation}
  M_i=\frac12(J_i+K_i),\quad N_i=\frac12(J_i-K_i),
  \label{MNdef}
\end{equation}
which furnish two independent SU(2) subalgebras:
\begin{equation}
  [M_i,M_j]=-\epsilon_{ijk}M_k\quad [N_i,N_j]=-\epsilon_{ijk}N_k,\quad
  [M_i,N_j]=0.
  \label{MNalgebra}
\end{equation}
To interpret these, we may note that a representation of $\{M_i,N_i\}$
acting on four-dimensional vectors can be given in terms of the {\it
  't Hooft symbols}
\begin{align}
  \eta^a_{\mu\nu} &= \tensor{\epsilon}{^a_{\mu\nu 4}} + \delta^a_{\mu}\delta_{\nu 4} - \delta^a_{\nu}\delta_{\mu 4}, \\
  \bar{\eta}^a_{\mu\nu} &= \tensor{\epsilon}{^a_{\mu\nu 4}} - \delta^a_{\mu}\delta_{\nu 4} + \delta^a_{\nu}\delta_{\mu 4},
  \label{etadefs}
\end{align}
satisfying (in matrix notation)
\begin{equation}
  [\eta^a,\eta^b]=-2\epsilon^{abc}\eta^c,\quad [\bar{\eta}^a,\bar{\eta}^b]
  =-2\epsilon^{abc}\bar{\eta}^c,\quad [\eta^a,\bar{\eta}^b]=0,
  \label{thooftalg}
\end{equation}
such that one may set
\begin{equation}
  M_i\rightarrow \frac12 \bar{\eta}^i_{\mu\nu},\quad N_i\rightarrow \frac12
  \eta^i_{\mu\nu}.
  \label{MNeta}
\end{equation}
From eq.~(\ref{etadefs}), one may verify that the 't Hooft symbols
satisfy
\begin{equation}\label{etaDuality}
  \eta^a_{\mu\nu} = \frac{1}{2}\epsilon_{\mu\nu\rho\sigma}\eta^a_{\rho\sigma}, \qquad
  \bar{\eta}^a_{\mu\nu} = -\frac{1}{2}\epsilon_{\mu\nu\rho\sigma}\bar{\eta}^a_{\rho\sigma},
\end{equation}
and thus are self-dual and anti-self-dual respectively. Their explicit
matrix representation, from eq.~(\ref{etadefs}) is
\begin{align}
\eta^1_{\mu\nu}&=&\left(\begin{array}{cccc}
0 & 0 & 0 & 1\\
0 & 0 & 1 & 0\\
0 & -1 & 0 & 0\\
-1 & 0 & 0 & 0
\end{array}\right),\quad
\eta^2_{\mu\nu}&=&\left(\begin{array}{cccc}
0 & 0 & -1 & 0\\
0 & 0 & 0 & 1\\
1 & 0 & 0 & 0\\
0 & -1 & 0 & 0
\end{array}\right),\quad
\eta^3_{\mu\nu}&=&\left(\begin{array}{cccc}
0 & 1 & 0 & 0\\
-1 & 0 & 1 & 0\\
0 & 0 & 0 & 1\\
0 & 0 & -1 & 0
\end{array}\right);\notag\\
\bar{\eta}^1_{\mu\nu}&=&\left(\begin{array}{cccc}
0 & 0 & 0 & -1\\
0 & 0 & 1 & 0\\
0 & -1 & 0 & 0\\
1 & 0 & 0 & 0
\end{array}\right),\quad
\bar{\eta}^2_{\mu\nu}&=&\left(\begin{array}{cccc}
0 & 0 & -1 & 0\\
0 & 0 & 0 & -1\\
1 & 0 & 0 & 0\\
0 & 1 & 0 & 0
\end{array}\right),\quad
\bar{\eta}^3_{\mu\nu}&=&\left(\begin{array}{cccc}
0 & 1 & 0 & 0\\
-1 & 0 & 1 & 0\\
0 & 0 & 0 & -1\\
0 & 0 & 1 & 0
\end{array}\right).
\label{thooftmatrices}
\end{align}
Further useful properties of these symbols, that we use throughout the
paper, are
\begin{align}
  \bar{\eta}^a_{\mu\nu}\,\bar{\eta}^a_{\rho\sigma}
  &=\delta_{\mu\rho}\,\delta_{\nu\sigma}-\delta_{\mu\sigma}\,\delta_{\nu\rho}
  -\epsilon_{\mu\nu\rho\sigma};\notag\\
  \bar{\eta}^a_{\mu\rho}\,\bar{\eta}^b_{\mu\sigma}&=\delta^{ab}\,
  \delta_{\rho\sigma}+\epsilon^{abc}\bar{\eta}^c_{\rho\sigma};\notag\\
  \epsilon^{abc}\,\bar{\eta}^b_{\mu\nu}\,\bar{\eta}^c_{\rho\sigma}&=
  \delta_{\mu\rho}\bar{\eta}^a_{\nu\sigma}
  +\delta_{\nu\sigma}\bar{\eta}^a_{\mu\rho}
  -\delta_{\mu\sigma}\bar{\eta}^a_{\nu\rho}
  -\delta_{\nu\rho}\bar{\eta}^a_{\mu\sigma}\notag\\
  \bar{\eta}^a_{\mu\nu}\bar{\eta}^b_{\mu\nu}&=4\delta^{ab}.
  \label{etaids}
\end{align}
These and additional identities may be found in e.g. appendix B of
ref.~\cite{Vandoren:2008xg}.\\

Thus far, the 't Hooft symbols have been described as representing
particular rotations in spacetime. However, they have a second
use in the study of non-abelian instantons, namely in embedding vector
fields into non-abelian gauge groups. Considering the case of pure
SU(2) Yang-Mills theory, this will have a matrix-valued gauge field
${\bf A}_\mu=A_\mu^a{\bf T}^a$, where $\{{\bf T}^a\}$ are the
generators of the gauge group. Given a vector field $V_\mu$, a general
procedure for turning this into a non-abelian field with components
$A_\mu^a$ is the {\it 't Hooft ansatz} of eq.~(\ref{thooft}). In this
equation, the upper index on the 't Hooft symbol is to be interpreted
as an adjoint index associated with the gauge group, rather than
labelling a given rotation generator. The 't
Hooft ansatz thus represents a map from a selected SU(2) subalgebra of $\text{SO}(4)$, to the SU(2) colour algebra. Embedding
vector fields into higher gauge groups is also possible, given that
the latter will contain SU(2) subgroups.


\section*{Acknowledgments}

We thank Silvia Nagy for many useful discussions, and collaboration on
related topics. We are also grateful to the participants of the
Amplitudes meeting at the Higgs Centre for Theoretical Physics for
useful feedback and advice. This work has been supported by the UK
Science and Technology Facilities Council (STFC) Consolidated Grant
ST/P000754/1 ``String theory, gauge theory and duality'', and by the
European Union Horizon 2020 research and innovation programme under
the Marie Sk\l{}odowska-Curie grant agreement No. 764850
``SAGEX''. KAW is supported by a studentship from the UK Engineering
and Physical Sciences Research Council (EPSRC).

\bibliography{refs}
\end{document}